%% file: cottensong2016arXiv.tex
\documentclass[preprint2]{aastex}
\usepackage{verbatim}
\usepackage{graphicx}
\usepackage{longtable}
\usepackage{lmodern} 
\usepackage{lscape}

\shorttitle{Nearby Infrared Excess Stars}
\shortauthors{Cotten \& Song}

\makeatother

\begin{document}

\title{A Comprehensive Census of Nearby Infrared Excess Stars}

\author{Tara H. Cotten}
\affil{Department of Physics and Astronomy, University of Georgia, Athens, GA 30602}
\email{tara@physast.uga.edu}

\author{Inseok Song}
\affil{Department of Physics and Astronomy, University of Georgia, Athens, GA 30602}
\email{song@physast.uga.edu}

\begin{abstract}
The conclusion of the WISE mission presents an opportune time to summarize the history of using excess emission in the infrared as a tracer of circumstellar material and exploit all available data for future missions such as JWST. We have compiled a catalog of infrared excess stars from peer-reviewed articles and perform an extensive search for new infrared excess stars by cross-correlating the Tycho-2 and AllWISE catalogs. We define a significance of excess in four spectral type divisions and select stars showing greater than either 3$\sigma$ or 5$\sigma$ significance of excess in the mid- and far-infrared. Through procedures including SED fitting and various image analyses, each potential excess source was rigorously vetted to eliminate false-positives. The infrared excess stars from the literature and the new stars found through the Tycho-2 and AllWISE cross-correlation produced nearly 500 {`Prime' infrared excess} stars and $\geq$1200 `Reserved' stars. The main catalog of {infrared excess stars} are nearby, bright, and either demonstrate excess in more than one passband or have infrared spectroscopy confirming the infrared excess. {This study identifies stars that display a spectral energy distribution suggestive of a secondary or post-protoplanetary generation of dust and they are ideal targets for future optical and infrared imaging observations.} {The final catalogs of stars} summarizes the past work using infrared excess to detect {dust disks} and with the most extensive compilation of infrared excess stars ($\sim$ 1750) to date, we investigate various relationships among stellar and disk parameters.
\end{abstract}

\section{Introduction}
{
Excess emission in the infrared (IR excess, hereafter) provides a useful tracer of the dust in a 
circumstellar disk due to the process by which the dust grains are heated 
by the starlight and reemit at longer wavelengths. 
Identifying each distinct evolutionary phase can be challenging 
but possible since the shape of the IR excess 
depends on the size, temperature, and composition of the emitting dust grains.
While it is understood that as the protoplanetary disk evolves, the  
gas and dust is cleared from the inner region closest to the star 
and then from the outer regions \citep{wyatt2015}, 
the discovery of the first debris disk around Vega \citep{aumann1984} using the 
\emph{InfraRed Astronomical Satellite} (IRAS; \citealt{beichman1988}), 
provided evidence for a secondary generation of dust. 
The secondary origin of this dust around a mature stellar system 
must be the result of the collisional grinding of planetesimals, comets, and asteroids \citep{kenyon2008}. 
The information gathered from IR excess stars provides a link to the formation 
and evolution of exoplanets \citep{wyatt2008}.} 

Many studies (\citealt{beichman2006b}, \citealt{bryden2006a}, \citealt{su2006}, \citealt{gautier2007}, \citealt{moromartin2007}, \citealt{rhee2007}, \citealt{hillenbrand2008}, \citealt{trilling2008}, \citealt{carpenter2009}, \citealt{greaves2009}, \citealt{morales2012}, \citealt{bulger2013}, \citealt{eiroa2013}, \citealt{patel2014})
have confirmed the use of excess emission in the infrared as an indicator
for circumstellar dust. 
Table \ref{irscopes} shows
the notable infrared surveys that were developed to improve the sensitivity of IRAS 
and were used in the detection of debris disk stars. 
Circumstellar material in locations analogous to our 
Kuiper and Asteroid belts can be detected by excess emission at $\geq$10$\mu m$ wavelengths and  
the past few decades has allowed for the exploration of disk properties 
such as dust temperature ranging from very cold ($\sim10$ K) to warm ($\sim500$ K). 
At the current time, all major mid- to far- infrared space missions have finished operations, including the most
recent {\emph{Herschel Space Observatory}} (\citealt{pilbratt2010}) and the \emph{Wide-Field Infrared Survey Explorer} 
(WISE; \citealt{wright2010}). Studies have shifted 
to analyze {data available through archives}. 
With the advent of new infrared missions still a couple years
in the future (e.g. {\emph{James Webb Space Telescope, JWST}; \citealt{gardner2006}}),  
{we are at a unique time in which we can devote
adequate effort to thoroughly characterize known IR excess stars in the solar neighborhood.}

The use of IR excess has generated hundreds of publications as well
as thousands of claimed IR excess stars.
Focusing specifically on sources that attest to dust undergoing recent collisional activity, as in the case of debris disks, 
each of these studies presented various source selection and infrared excess criteria.
However, many problematic IR excess candidate stars present conflicting evidence based on these different criteria. 
Therefore, a vetted list of IR excess stars is needed to maximize the scientific return of imminent next generation missions 
(JWST and WFIRST; \emph{Wide-Field InfraRed Survey Telescope}) and currently ongoing missions 
with new extreme adaptive optics instruments such as GPI (\emph{Gemini Planet Imager}, \citealt{macintosh2006}) and 
SPHERE (\emph{Spectro-Polarimetric High-contrast Exoplanet REsearch}, \citealt{beuzit2008}).
Yet, we need to unify the search for IR excess stars through simultaneous examination of both previous 
and new findings of IR excess. 

The most comprehensive catalog of nearby IR excess stars is created here through the combination 
of 1) a literature search for mostly far-IR excess stars discovered
with the \emph{Spitzer Space Telescope}, \emph{Herschel Space Observatory,}
and IRAS (Section \ref{litsearch}) and 2) a new search for mid-IR excess stars 
using the all-sky \emph{Wide-Field Infrared Survey Explorer }catalog (AllWISE, \citealt{wright2010}; Section \ref{cross-match}).
The paper begins with a description of our literature
search of over 200 published articles that present many IR excess and
debris disk stars and we describe our reanalysis of the claimed IR excess.
Section \ref{cross-match} recounts our new investigation into
IR excess stars with warm dust radiating at $\geq$10$\mu$m through the cross-correlation between
the Tycho-2 catalog (\citealt{hog2000}) and the AllWISE catalog,
providing significant additions to historical infrared studies. We
compare our Tycho-2/AllWISE cross-match to similar warm dust studies
performed recently in Section ~\ref{comparison}. Characteristics of
the final IR excess catalog is provided in Sections \ref{samplechar} and \ref{discuss},
followed by our conclusions and future work described in Section \ref{conclude}.

\section{Literature Search}\label{litsearch}

Following \citet{aumann1984}, many studies reported new discoveries of IR excess stars. 
To summarize known IR excess stars from the literature, we select a few 
pivotal investigations involving IRAS and \emph{Spitzer}
as well as a review article: \citet{rieke2005}, \citet{rhee2007},
and \citet{wyatt2008}. Besides the lists of authentic excess stars
in each of these reports, citations included in these articles  
provide the basis of our literature search. We meticulously
comb over 230 articles (displayed in Table \ref{refs}) that cite at least one of these three 
pivotal papers for stars claimed to have IR excess and compile a database of these objects. The collection
of previous publications excludes any searches for IR excess candidates
that singularly used the \emph{Wide-Field Infrared Survey Explorer} (WISE), 
since this was the intent of our new infrared search presented later
in Section ~\ref{cross-match}.

In our creation of the main IR excess star catalog, we focus on
identifying nearby, main sequence stars with post-protoplanetary disks. 
However, without available age information, we select disks through a characteristic inspection of the shape 
of the spectral energy distribution. 
We assume that stars having photospheric AllWISE fluxes at W1 (3.5$\mu$m) and W2 (4.6$\mu$m) bands 
with clear excess emission at  mid- to far-IR are sources involved in secondary dust generation. 
We reserve a discussion of the ages of these systems for a forthcoming paper using optical spectroscopy to characterize our primary target IR excess stars. 
Using these criteria, it is reasonable to assume that we will be missing the youngest disk counterparts including many T Tauri and Herbig Ae/Be disks.
Furthermore, we avoid publications that sought to find circumstellar disks around very distant stars ($>$ 500 pc; e.g. \citealt{cloutier2014}) 
or white dwarfs (e.g. \citealt{barber2012}). 
For example, we keep only stars from \citet{luhman2012} that can be matched best to an `older', inner cleared disk with no excess emission at W1 or W2.
Moreover, some studies provide lists of rejected sources (\citealt{rhee2007}, 
\citealt{ballering2013}) that enabled us to avoid stars 
confirmed to be non-excess.

To investigate the nature of apparent IR excess emission, we create a spectral
energy distribution (SED) of all compiled candidate IR excess stars and perform several procedures to reduce including any false-positives.
The SED displays flux density versus wavelength from measured photometry which is then fit to a Phoenix NextGen main sequence stellar model \citep{hauschildt1999}.  
For more details regarding the SED models and fitting, refer to \citet{rhee2007}.
Roughly 820 stars were compiled and assessed from the literature search. 
However, since previous publications include evolved stars whose IR excess mechanism is
different from that of main sequence stars, we checked all IR excess candidate stars
using SIMBAD and remove 20 stars with luminosity classes of I, II or III. 
Lastly, two key procedures for eliminating false-positive excess stars include the visual inspection of the SED  
and AllWISE images to remove possible contaminated sources (explained in full detail in Section ~\ref{visualcut}). 
We use the VizieR database to gather additional photometry from optical to far-IR wavelengths if available.
The multitude of photometric measurements ensures a reliable SED fit to the stellar photosphere 
and enables us to quantify the number of passbands which display IR excess (`Num\_Excess').  
Section \ref{distcut} has more details regarding the parameter `Num\_Excess'.  
We remove 65 stars having photospheric flux at far-IR wavelengths (i.e. showing no IR excess).

Among IR excess publications, \citet{chen2014} is notable because they
provided Spitzer IRS spectra of over 300 stars. When available, they also reported
MIPS 24 and 70 $\mu$m measurements. Because the evidence of excess determined from infrared spectroscopy 
aids in the reliability of IR excess from photometry alone, we
retain most stars from \citet{chen2014} unless the IRS spectra is consistent with
the photosphere.

To create the most useful catalog of IR excess stars for more efficient future
follow-up observations, we implement additional restrictions for a star to be included in the
``Prime'' table: (1) AllWISE W3 (12 $\mu$m) or W4 (22 $\mu$m) flux being greater than 10 mJy, (2) distance
within 120 pc and (3) either multiple passbands demonstrating IR excess (`Num\_Excess'$>$1) or 
1 passband of IR excess and IRS spectroscopic confirmation of the photometric excess. 
These requirements ensure that stars in our final catalog are bright
enough to be fully characterized. The distance restriction is sometimes
relaxed to include few interesting stars with corroborating IRS spectra, however, no star beyond
150 pc was included. Our choice of distance cut, 120 pc, is mainly to remain inside of
the local bubble so that we can ignore interstellar reddening and nearby star-forming regions. 

After all removals and restrictions, the literature sample of ``Prime'' IR excess
stars (Table \ref{highfidelity}) contains $\sim$430 unique targets. 
Distant, faint, and marginal excess candidates are maintained in our ``Reserved'' star catalog 
(Table \ref{reserved}, an additional $\sim$300 stars).
Hereafter, we shall refer to this list as the ``literature IR excess stars''.
The catalog of literature IR excess stars and information regarding
the star and disk parameters for this sample of stars is included
in Tables \ref{highfidelity} and \ref{reserved}. 
Each table's details are explained fully in Section \ref{distcut}.

\section{Tycho-2 \& AllWISE}\label{cross-match}

\subsection{Total Proper Motion and Cross-Correlation}\label{TPM}

Analysis of photometric excess based on stellar SEDs requires precise optical
photometry in order to constrain the spectral shape (i.e., the stellar
effective temperature). 
Thus, we cross-correlate a large optical survey (the Tycho-2 Catalog of the 
2.5 Million Brightest Stars as released in 2000; \citealt{hog2000}) and the 
most comprehensive all-sky mid-infrared survey (the AllWISE all-sky catalog; 
\citealt{cutri2013}) to create a massive list of sources.
Lacking accurate parallax measurements for most Tycho-2
stars, we implement a restriction on the proper motion magnitude defined by: 

\begin{center}
\begin{equation}
\mu_{total}=\sqrt{(\mu_{\alpha})^{2}+(\mu_{\delta})^{2}}\geq25.0\ (mas/yr),
\end{equation}

\par\end{center}
as a proxy for distance corresponding to stars within 200 pc. 
The stars within 100 pc that also have total proper motions greater than
25 mas/yr recovers 91$\%$. 
The proper motion criterion solely applied to the Tycho-2 catalog assembles 515,518 stars.

We perform a cross-match of the proper motion selected Tycho-2 sample
and the AllWISE survey comparing the catalog positions. 
Considering the twenty year baseline between these two catalogs implies an object
with a large proper motion ($\gtrsim$ 0.1\arcsec{}/yr) would be displaced
by 2.0\arcsec{}, however, we execute a proper motion correction to Tycho-2
positions using the Tycho-2 proper motions in order to mitigate this effect. 
We select sources with a 5.0\arcsec{} match radius
between AllWISE and Tycho-2 sources and the cross-match 
returns 99.6$\%$ of the Tycho-2 sample. 
The cross-correlated sample contains 513,478 objects.

\subsection{IR Excess Selection Procedures}\label{candidates}

\begin{center}
\begin{figure}
\begin{centering}
\includegraphics[width=6in]{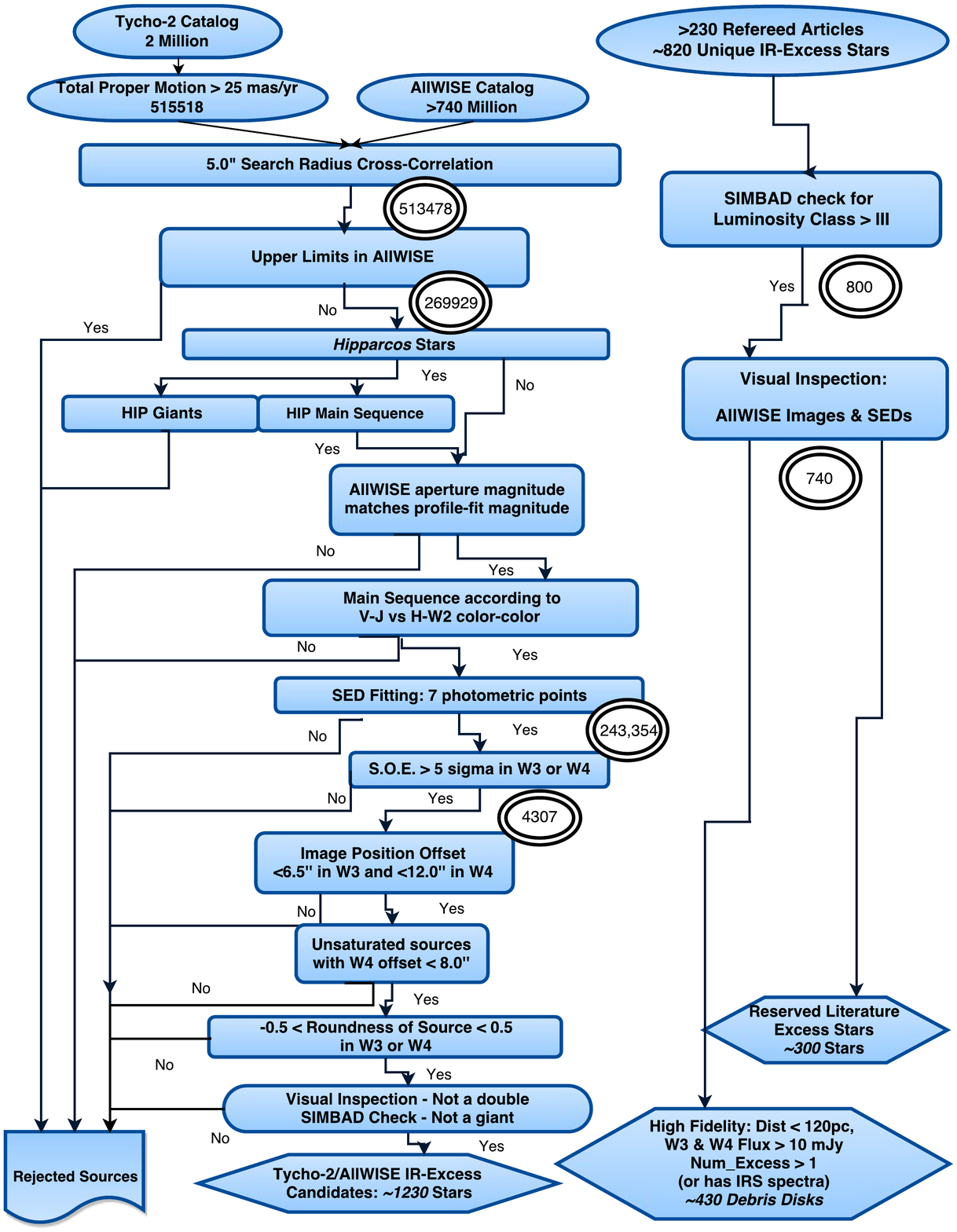}
\end{centering}
\caption{Flowchart description of the initial
sample selection used to collect the Tycho-2/AllWISE cross-correlated
IR excess candidates and a summary of literature IR excess stars selection. }
\label{flowchart}
\end{figure}
\end{center}

This section provides our algorithm to reduce the starting sample from over
500,000 stars to a reliable sample of main-sequence, IR excess candidate stars.
A number of criteria are used to remove evolved stars, poor photometric
quality data, and false positives. A summary of the procedure can be found in the flowchart
shown in Figure \ref{flowchart}.

\subsubsection{Giants}\label{giants}

A non-negligible fraction of the cross-correlated sample will inherently
be evolved stars that are not discernible from main sequence stars.
Although there are some interesting exceptions of giants with IR excess (i.e.
Phoenix giants, see \citealt{melis2009}), we will focus only on main
sequence stars where IR excess points to circumstellar material
suggestive of planetary relevance. 

\begin{center}
\begin{figure}
\begin{centering}
\includegraphics[width=6in]{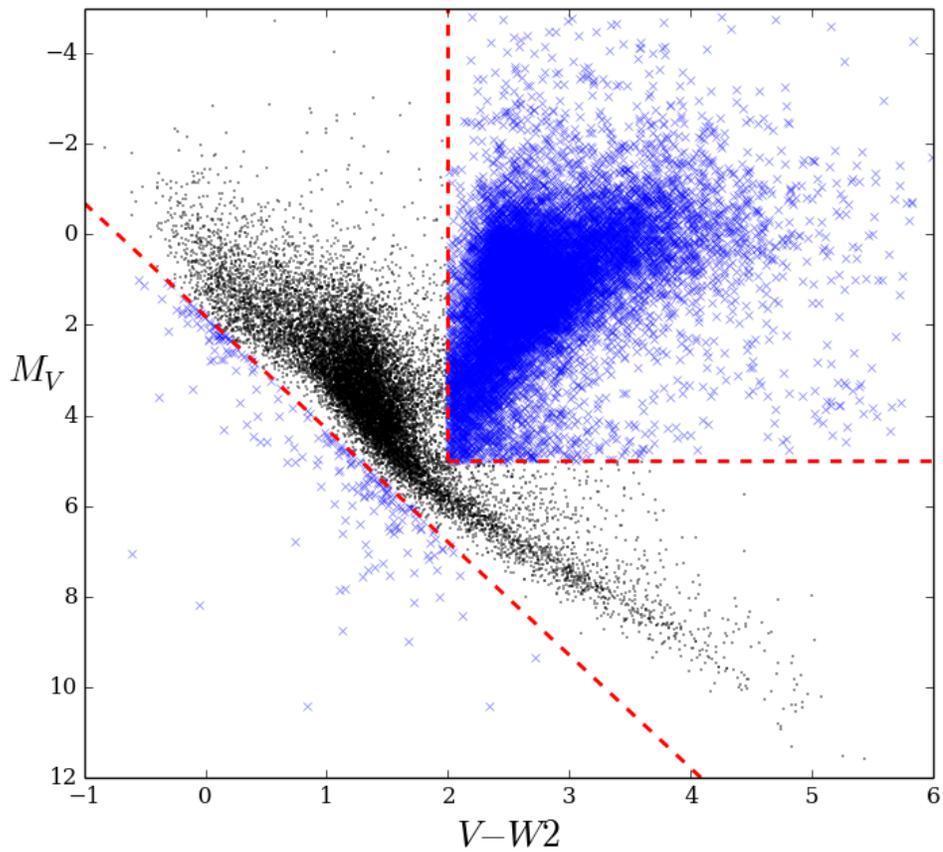}
\protect\caption{Color-magnitude diagram for the Tycho-2/AllWISE
stars with \emph{Hipparcos} parallax information.
The dashed lines designate the cuts for removing giants and white dwarfs. We have
selected this line to ensure we do not lose many main sequence dwarfs.
The blue `x' symbols constitute over 15000 stars rejected from the cross-correlated
sample for being giants or white dwarfs.}
\label{hip_cmd}
\end{centering}
\end{figure}
\end{center}

To gain insight into the contamination fraction of giant stars
in our sample, we cross-match our sample of 513,478 Tycho-2/AllWISE 
sources with the \emph{Hipparcos} catalog \citep{leeuwen2007}.
A 5.0\arcsec{} search radius ensures the closest match 
with the \emph{Hipparcos} catalog and returns 54,016 stars with measured parallax.
A similar color-magnitude diagram (CMD) method of excluding evolved stars using the \emph{Hipparcos}
catalog was performed by \citet{rhee2007} and more recently by \citet{patel2014}.
We convert the Tycho-2 B$_{T}$ and V$_{T}$ magnitudes 
to the Johnson system using correction factors \citep{bessell2000}.
We choose (V-W2) for the color of our CMD since our sources have 
AllWISE data and this color provides the longest, useful color baseline. 
Figure \ref{hip_cmd} displays (V-W2) color versus the absolute visual magnitude for
the 54,016 Tycho-2/AllWISE/\emph{Hipparcos} sources and the well-structured evolutionary separations 
between the white dwarfs, main sequence, and giant branch stars. 
Our choice of excluding giants and white dwarfs is shown by the dashed red lines
and blue `X's in Figure \ref{hip_cmd} defined by:
\begin{equation}
(V-W2)>2.0\ and\ M_{V}\leq5.0
\end{equation}
\begin{equation}
M_{V}\geq2.5\times(V-W2)+1.8.
\end{equation}
This procedure identifies 15,071 stars as giants and 581 white dwarf stars among 54,016 Tycho-2/AllWISE/\emph{Hipparcos} stars.
The fraction of evolved stars from a sample of about 54,000 is about 30\% and we expect this same contamination rate of the 
Tycho-2/AllWISE sample that lack \emph{Hipparcos} data.  
We anticipate the majority of remaining giant stars to belong to K and M spectral types since early type giants are rare
due to their short lifetimes.

\begin{center}
\begin{figure}
\begin{centering}
\includegraphics[width=6in]{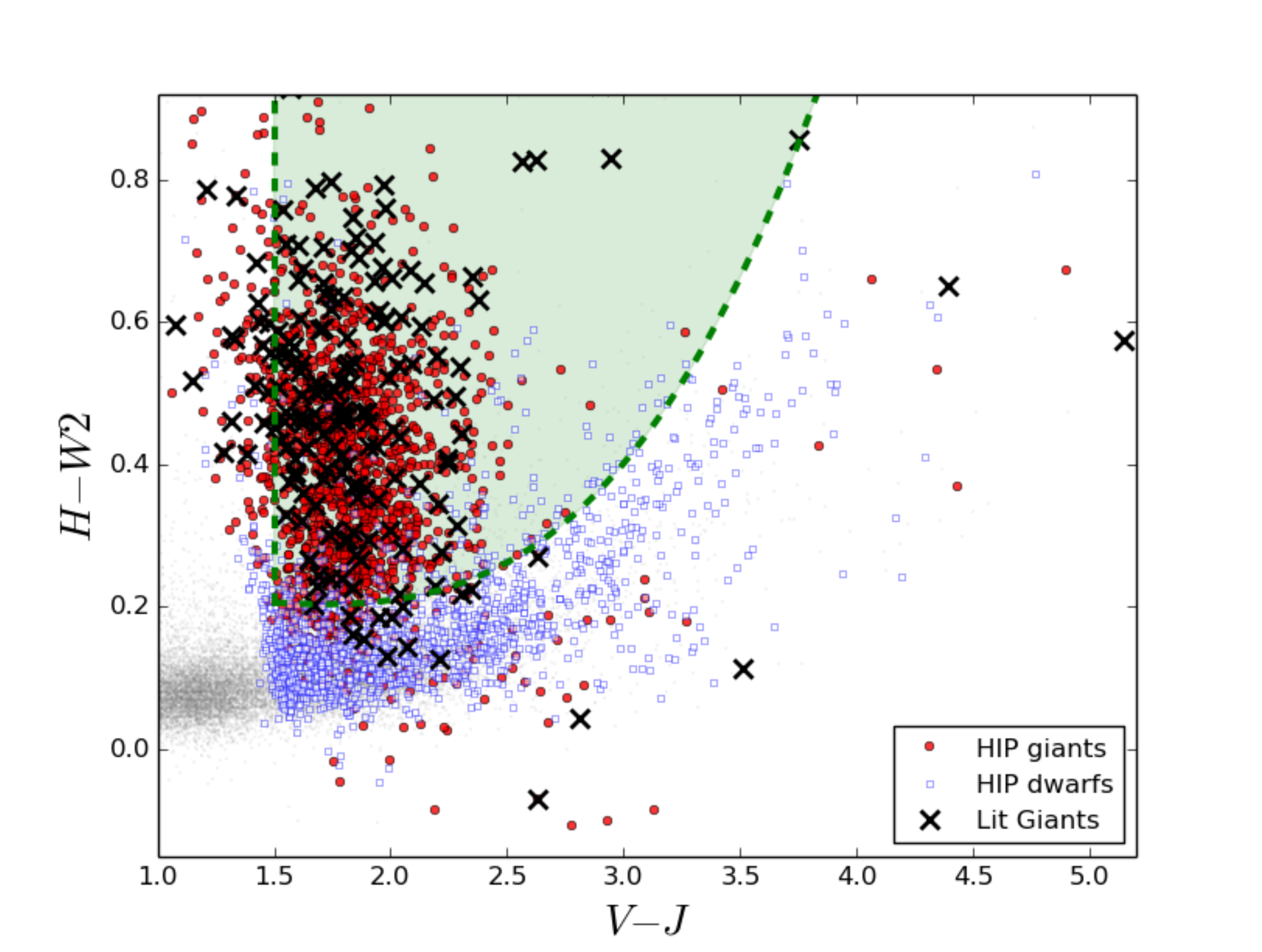}
\protect\caption{Color-color diagram used to distinguish
giant stars. The green dashed curves and the enclosed shaded region 
displays the cuts used in removing additional late-type 
evolved stars from our main sequence sample (see
Section ~\ref{giants} for the functional form of the curved line).
The small grey dots are the remaining sample of $>200,000$ Tycho-2
stars. The red circles are the \emph{Hipparcos }giants
removed from the sample using the CMD (see Figure \ref{hip_cmd}), while the blue, unfilled squares are
the \emph{Hipparcos} main sequence sample selected
in the CMD that have well-measured parallax (error $<$10\%). 
The large `X' symbols are giants selected from SIMBAD which have a luminosity class of I, II, or III.}
\label{giantcolor}
\end{centering}
\end{figure}
\end{center}

Since the CMD method requires 
\emph{Hipparcos} data, we also investigate a method to remove evolved
stars based on various color-color diagrams. 
\citet{bessell1988} demonstrated using color-color diagrams to identify 
the divergence of late-type main sequence and giant tracks that appear to diverge at approximately early M type.
For our study, we want to identify colors 
which are able to distinguish G and K type dwarfs from giants. 
We compare dwarf and giant model fluxes at various passbands to determine a useful color which bifurcates 
the evolved branch at spectral types earlier than M.
We find that the broadband colors of H - W2 versus V - J 
shows this distinction most clearly (Figure \ref{giantcolor}).
We define the conservative polynomial to remove
objects falling within the red dashed curves:
\begin{equation}
y=0.05x^{3}-0.19x^{2}+0.23x+0.13
\end{equation}
 where $x=$ (V-J) and $y=$ (H-W2). The selection using the 
dashed lines removes 77\% of the known \emph{Hipparcos} 
giant sample and 82\% of the literature giant sample. 
Using the V-J and H-W2 color-color cut, it demonstrates that about 80\% of known giants can be
flagged while only about 2\% of main sequence stars are lost. So, by applying this
color-color cut in combination with the expected contamination rate from the CMD, 
we expect that only $\sim$6\% of the final Tycho-2/AllWISE sample will be giants.

\subsubsection{AllWISE Photometry}

We compare the profile-fit and aperture derived photometry as an additional
analysis of the reliability of the AllWISE photometric measurements.
While the AllWISE explanatory supplement recommends using the profile-fit photometry to avoid the poor 
aperture photometry at the saturation limits (W1$>8.0$, W2$>7.0$,W3$>3.8$; \citealt{cutri2013}), we believe 
that unsaturated, well-behaving stars should have similar photometry derived from both methods. 
We remove sources for which this is not the case by fitting a standard Gaussian to the  
differences between the aperture and profile-fit magnitudes at each AllWISE passband 
and exclude stars having differences in photometry outside of 2.5 $\sigma$. 
Nearly 6000 stars had photometry that we would term unreliable. 
These sources tend to be contaminated by nearby ($\lesssim$12\arcsec{}) brighter
objects such as stars, galaxies, or nebulae or cases of strong cirrus contamination in either W3 or W4. 
The number of stars in our sample after removing the giants and poor photometry sources is 245,924. 

For accurate SED fitting, we remove over two thousand stars without a complete set of measurements 
from Tycho-2, 2MASS, and AllWISE. 
Some of the stars in our final sample may have poorer quality data according to the flags from the 
individual catalogs. However, we continue to use this data rather than disregard those magnitudes as it improves the SED 
fitting.  The only instance that photometric measurements do not improve the SED fit 
are upper limits and so we have eliminated stars with upper limits at these crucial wavelengths. 

\subsubsection{SED Fitting}\label{seds}

In our SED fitting, NextGen (PHOENIX code version 9.1, \citealt{hauschildt1999}) 
and Kurucz model atmospheres are fit against observed photometric 
measurements. Between two models, there is a systematic difference of about $\sim$120 K 
in the best fit stellar temperature. 
Because of the small difference between two models and to avoid mixing 
two models in our analysis, we decide to use only the NextGen models in our SED fitting.
The SED fitting algorithm converts all the photometric measurements
into flux and compares them to the grid of available
temperatures and selects the best agreement between model and data
using a $\chi^{2}$ minimization technique (Refer to \citealt{rhee2007} for complete details). 
After producing a good fit to the photosphere, the mid-infrared Rayleigh-Jeans tail provides a 
comparison for excess emission above the photosphere.

\subsubsection{Spurious AllWISE Saturation Correction}\label{saturation}

\begin{center}
\begin{figure}
\begin{centering}
\includegraphics[width=6in]{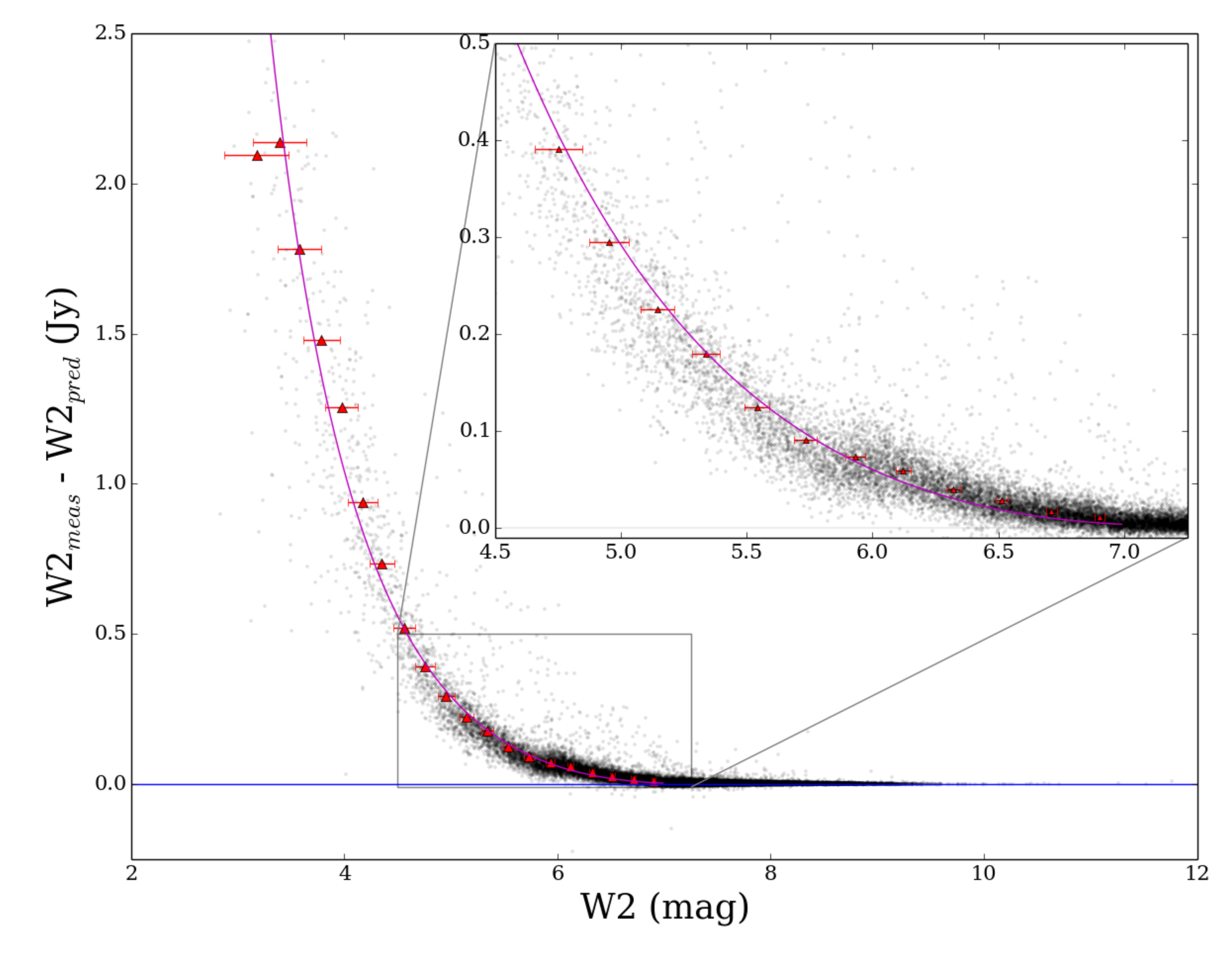}
\protect\caption{The function displayed is designed to correct the flux
over-estimation effect in AllWISE W2 passband. The data plotted
contains stars with effective temperatures from the SED fitting greater
than 6000K, binned by $\sim$0.2 magnitudes, and fit using a series of logarithms in the saturation
region. A solid (blue) horizontal line has been displayed at zero for reference. The correction function 
is of the form: $y(Jy) = 3.28 - 168.55\times\log(x+0.084)^{-1}+164.78\times\log(x+0.003)^{-1}$ and applies to 
W2 magnitudes less than 7.}
\label{sat_func}
\end{centering}
\end{figure}
\end{center}

Upon inspecting a number of SED fits using the procedure described
in Section \ref{seds}, we noticed an issue of overestimated fluxes 
in AllWISE measurements.
Sources brighter than 8.1, 6.7, 3.8, and -0.4 mag at W1 through W4, respectively, are saturated and thus, 
the fluxes are overestimated \citep{cutri2013}.
\citet{patel2014} presents a similar discussion but a different analysis.
The WISE team described this bias for W2 $<$6.5 mag and illustrated this finding
through their Figure 8 in Section VI.3.c.4 of \citet{cutri2012},
however, at the time of writing they did not offer a solution to the
nearly 0.5 mag over-estimate for the brightest objects. The bright
candidates in our sample demonstrate this spurious flux mainly at W2. 
Since this effect is not intrinsic to the object, 
we develop a correction to the AllWISE W2 flux using
over 26,000 early A and F stars ($T_{*,SED}\geq6000$ K) selected by their best fit stellar temperature from the SED. 
These earlier type stars have a smooth Rayleigh-Jeans
tail at every AllWISE passband while later type stars develop strong
carbon monoxide (CO) absorption features near W2 (4.6$\mu$m). 
Figure \ref{sat_func} displays the over-estimation of the flux density specifically for 
magnitudes brighter than 7 in W2.
The correction function involves a series of logarithms as described below:
\newline \begin{center} W2 $\leq$ 7.0 mag: \end{center}
\begin{equation}
y=3.28 - 168.55\log(x+0.084)^{-1} + 164.78\log(x+0.003)^{-1}
\end{equation}
where the value of y refers to the difference between the measured and predicted W2 flux 
in Jansky and the value of x refers to the AllWISE catalog magnitude. 

\subsubsection{Qualification of Excess}\label{irexcess}

Our infrared excess qualifications use the predicted flux values at the AllWISE passbands determined 
by the best fit SED after applying the saturation correction. 
We define the amount of IR excess in terms of a `Significance of Excess' as: 
\begin{center}
\begin{equation}
\mathrm{Significance\ of\ Excess\ }\equiv\frac{F_{AllWISE}-F_{predicted}}{\sqrt{(\sigma_{AllWISE}^2 + \sigma_{cal}^2)}}
\end{equation}
\par\end{center}
where $F_{AllWISE}$ is the measured flux at a given AllWISE passband
and $\sigma_{AllWISE}$ is the uncertainty in that measurement combined
with an absolute calibration uncertainty ($\sigma_{cal}$) defined by \citet{jarrett2011}
and \citet{cruzsaenz2014} to be 4.5\% in W3 and 5.7\% in W4. The
calibration uncertainty was derived by \citet{jarrett2011} through
comparison of AllWISE photometry and \emph{Spitzer} data for a set of standard stars.
$F_{predicted}$ represents the photospheric flux value
from the SED fit predicted at each AllWISE passband. 
As mentioned previously, 
we expect W1 and W2 to be consistent 
with the stellar photosphere and use W3 and W4 excess for our final IR excess candidates. 
In our analysis, we do not include a color correction of AllWISE measurements because the effects are small.

\begin{center}
\begin{figure}
\begin{centering}
\includegraphics[width=6in]{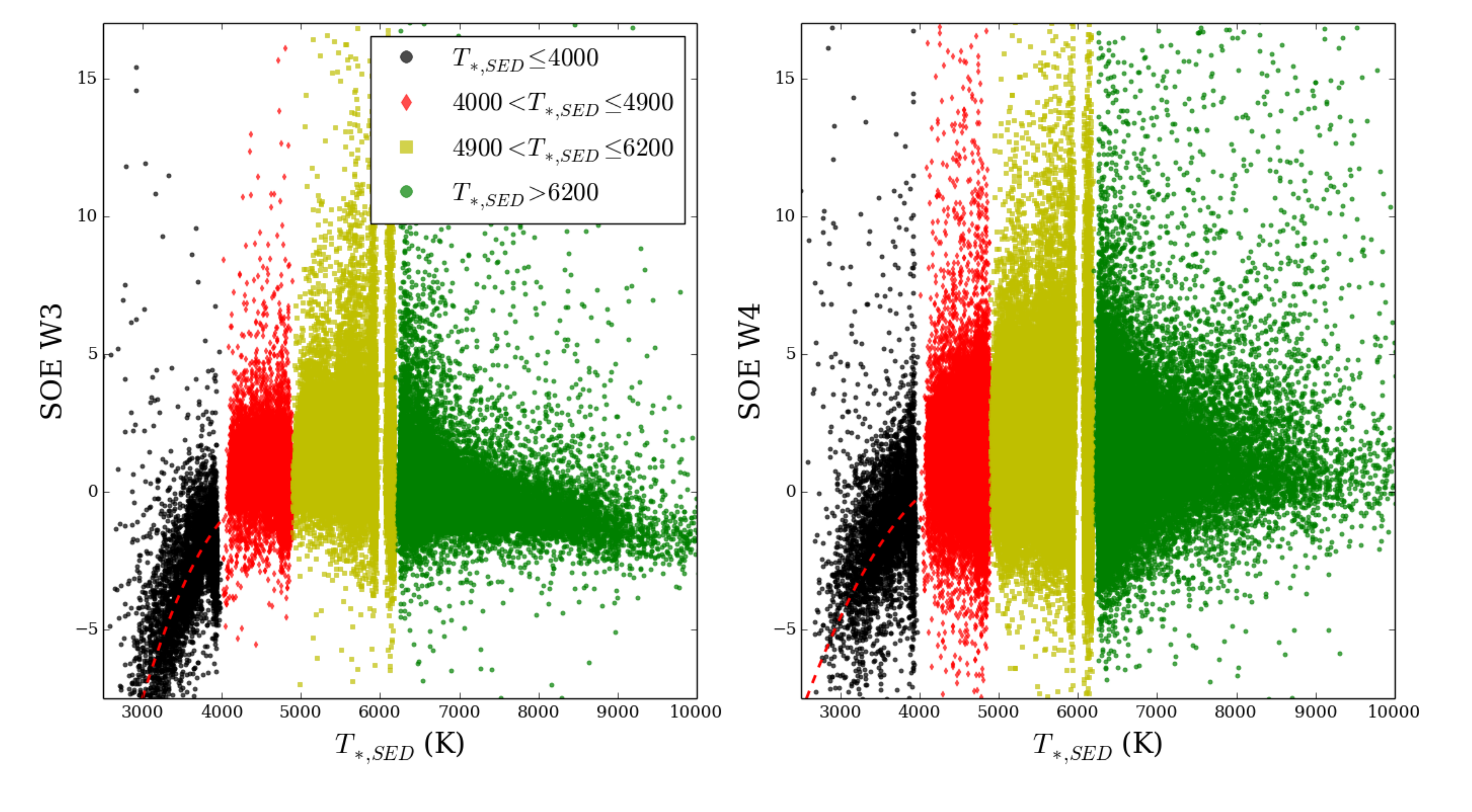}
\par\end{centering}
\protect\caption{\emph{Left:} Significance
of excess in AllWISE W3 flux for stars in the cross-correlated sample
(243,354) versus the best fit stellar temperature from the SED. 
The colors represent four different regions pointing out the trend towards `negative excess` of 
the coolest stars in our sample. This trend is seen for stars with SED temperatures less than 4000K 
and is fitted with a function shown as the dashed line (Refer to Section \ref{irexcess}: Equations 7 and 8 for the 
functional form of the corrections). 
\emph{Right:} Same as the left plot for AllWISE W4. }
\label{significance}
\end{figure}
\end{center}

Figure \ref{significance} displays the significance of excess versus stellar temperature 
for the 243,354 stars in our sample.
The figure shows a significant decreasing trend in the significance of excess with decreasing stellar 
temperature, in particular, for stars with $T_{*,SED}<4000$ K.
AllWISE W3 shows a steeper decline and we 
believe this is inherently due to the larger passband of W3 than W4. 
\citet{wright2010} mentions a color correction applied to the
flux in W3 would be larger than for any other passband and is exacerbated by the variability and activity 
found around nearby late-type stars. Further, \citet{wright2010}
describes the in-flight discrepancy found between red and blue sources that 
implies that the coolest stars will have a W3 flux that is measured
to be fainter than models (i.e. a negative significance of excess). 
To remove this effect, we fit a curve to this trend for stars with temperature
less than 4000K. The functional form of the curve for the W3 and W4
significance of excess (SOE) correction is:
\begin{equation}
W3:\ SOE = -2.9\times{}10^{-6}\times{}T_{*}^{2} + 0.03\times{}T_{*} - 62.22
\end{equation}

\begin{equation}
W4:\ SOE = -1.7\times{}10^{-6}\times{}T_{*}^{2} + 0.02\times{}T_{*} - 37.92
\end{equation}
where $T_{*}$ refers the the best fit stellar temperature from the SED. 
The stars in the other temperature regions do not
show any significant trend requiring a correction. 

\begin{center}
\begin{figure}
\begin{centering}
\includegraphics[width=6in]{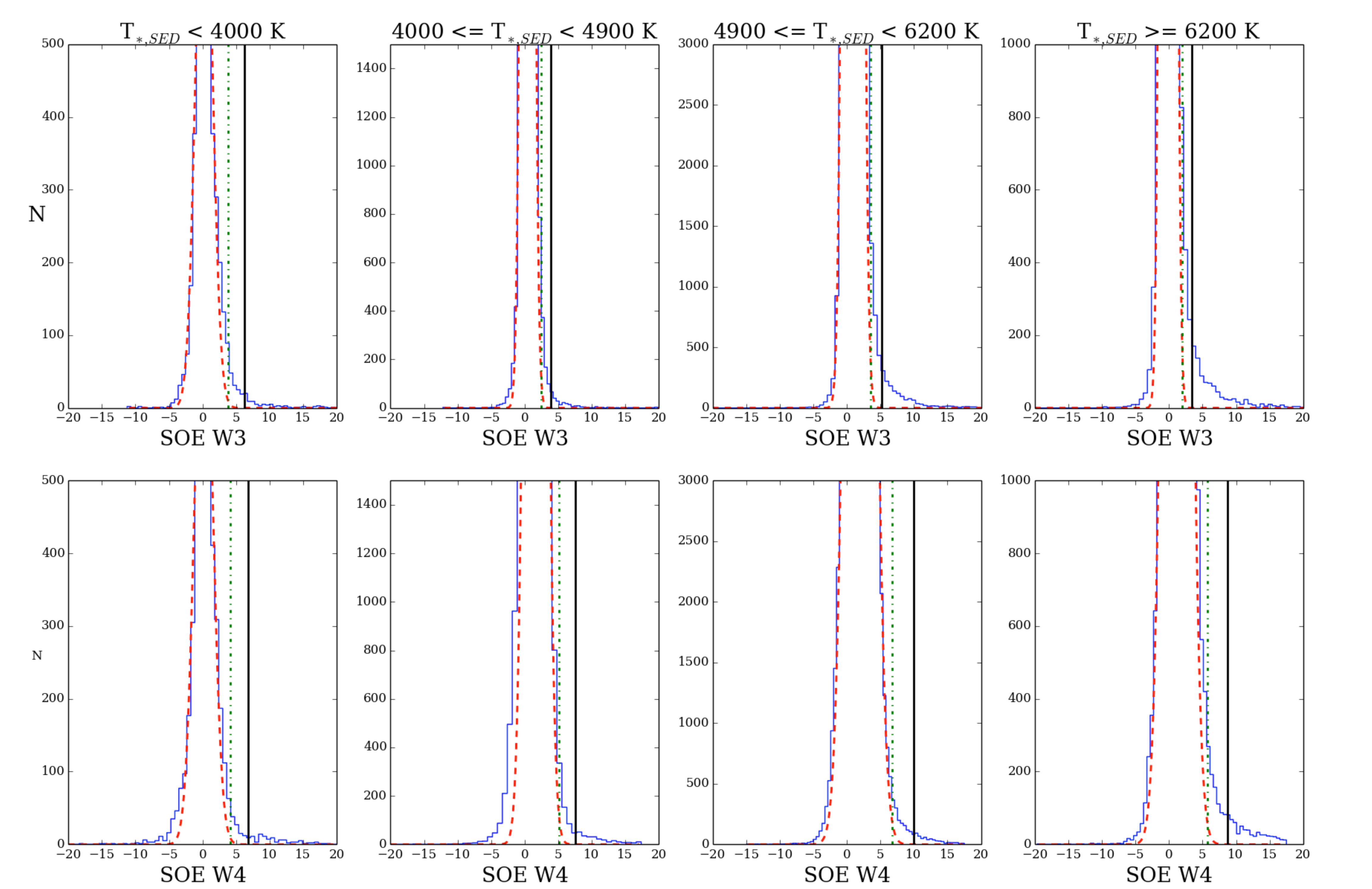}
\protect\caption{Significance
of excess for the temperature divisions shown in Figure \ref{significance}.
Each histogram portrays our significance of excess defined in Section \ref{irexcess} and 
is fitted with a Gaussian and the green (dot-dashed) and black (solid) vertical lines represent
the 3$\sigma$ and 5$\sigma$ selection criteria for excess stars, respectively. 
The histograms have been magnified to show the true distribution of stars within each temperature division.}
\label{hists}
\end{centering}
\end{figure}
\end{center}

After applying the correction to the stars with the coolest temperatures ($T_{*,SED}<4000$K), 
histograms of the significance of excess are displayed in
Figure \ref{hists} for each temperature division
for W3 and W4. Our selection of significant IR excess uses a Gaussian fit 
to each apparent population of non-excess stars shown by the red dashed curve.  
The mode of the significance of excess is offset from zero in the positive direction in many histograms, 
likely due to a combination of model uncertainties, Malmquist bias, and other unknown uncertainties.
We initially select the best excess candidates using the Gaussian fits and retain 
stars beyond the solid, vertical, black lines
representing 5$\sigma$ in W3 or W4 for each temperature division, however,  
recognizing that many past studies (\citealt{vican2014}, \citealt{patel2014}) 
isolated cases of marginal W4 excess which may in fact prove to be true detections of IR excess, 
we also include the sample of stars with W3 or W4 greater than 3$\sigma$ shown by the vertical dashed line in Figure \ref{hists}.  
This procedure identifies nearly 4300 stars with significant IR excess.

\subsubsection{Contamination Inspection}\label{visualcut}

Given the likelihood that many false-positive stars are due to source confusion
in AllWISE images due to a large mid-IR beam size, 
a series of quantitative image analyses are presented.  
These sources of contamination include cirrus or foreground infrared
sources, adjacent stronger IR sources affecting the AllWISE photometry,
background galaxies, nebulosity surrounding the source (see Pleiades
phenomenon \citealt{herbig2001}, \citealt{kalas2002}), and optical/near-infrared
binary stars which cannot be resolved with the beam size of WISE.
The goal of the image analysis is to eliminate the 
contamination issues in a quantitative fashion. 

The first methodology aims to compare the expected central, cross-correlation position to 
the position found through isolating the brightest central source in each AllWISE image. 
When a contamination source is present (at W1 and W2) adjacent to a candidate IR excess star and is unresolved
in the AllWISE image, the centroid position of the candidate star shifts. 
The second methodology is to analyze the isolated source's shape to
determine if the object is extended or noncircular. 
We use a criteria of roundness defined through
comparison of the bilateral symmetry of each source determined by
fitting a two dimensional gaussian to the source point-spread function
defined similar to: 
\begin{equation}
Roundness\propto\frac{(\sigma_{x}-\sigma_{y})}{\frac{(\sigma_{x}+\sigma_{y})}{2}}
\end{equation}
where $\sigma_{x}$ and $\sigma_{y}$ are the standard deviations of those gaussians.
A roundness criteria of zero would appear
circular while a roundness of -1.0 or 1.0 would be noticeably elliptical. 
Further, since we expect W3 and/or W4 excess to be 
from the dust grains, the disks should 
be unresolved at the W3 and W4 passbands (except for the nearest stars), hence, their roundness 
values should be close to zero. 

\begin{center}
\begin{figure}
\includegraphics[width=6in]{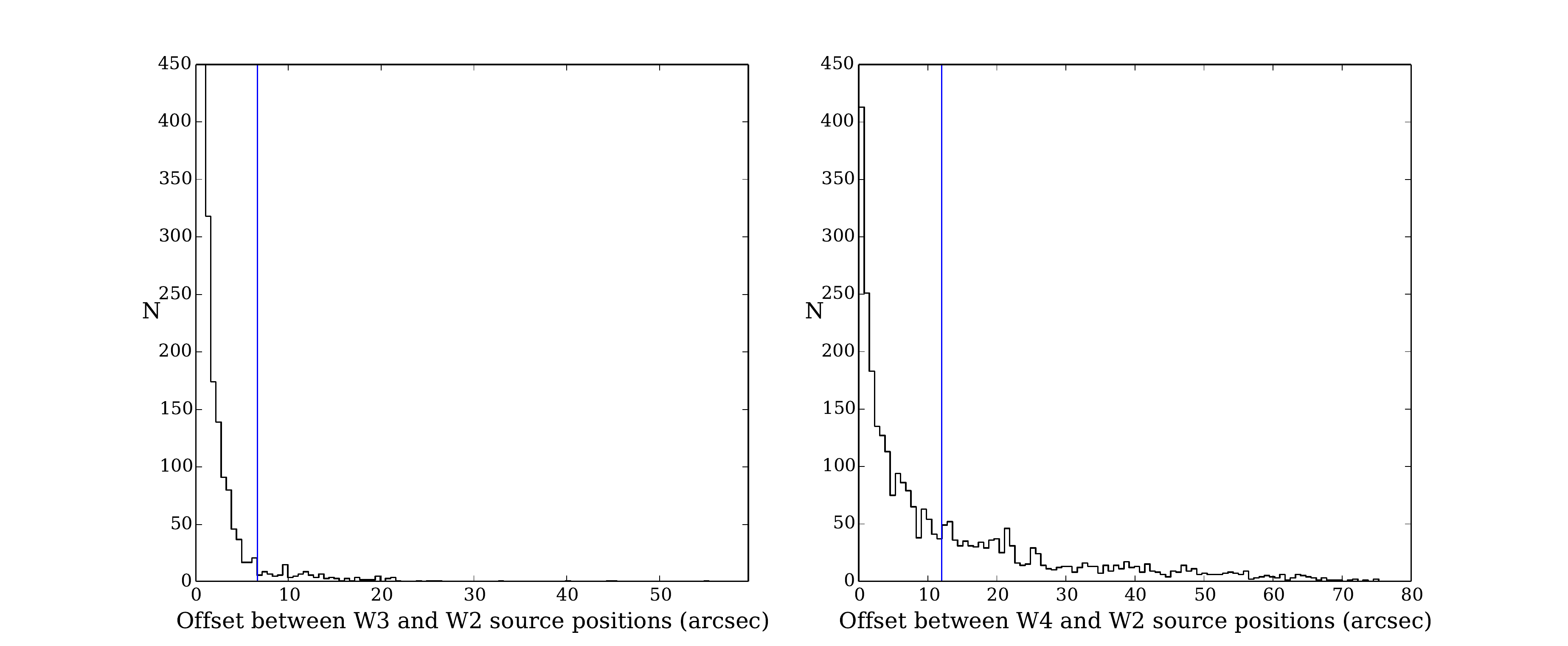}
\protect\caption{Offset in the main source position between AllWISE
W2 and W3 (\emph{Left}) and W2 and W4 (\emph{Right}).
The difference between W2 and W3 do not show a significant
variation from which to distinguish contamination. The difference
between W2 and W4 shows a larger spread and can easily distinguish
cases of image contamination. We have initially removed stars further separated
than 6.7\arcsec{} in W3 or 12.0\arcsec{} in W4 using the resolution of WISE as a
cutoff. The vertical lines indicate this location. (See text for more details.)}
\label{posoffset}
\end{figure}
\end{center}

\begin{center}
\begin{figure}
\centering
\begin{tabular}{ccc}
a. & \includegraphics[height=2.2in]{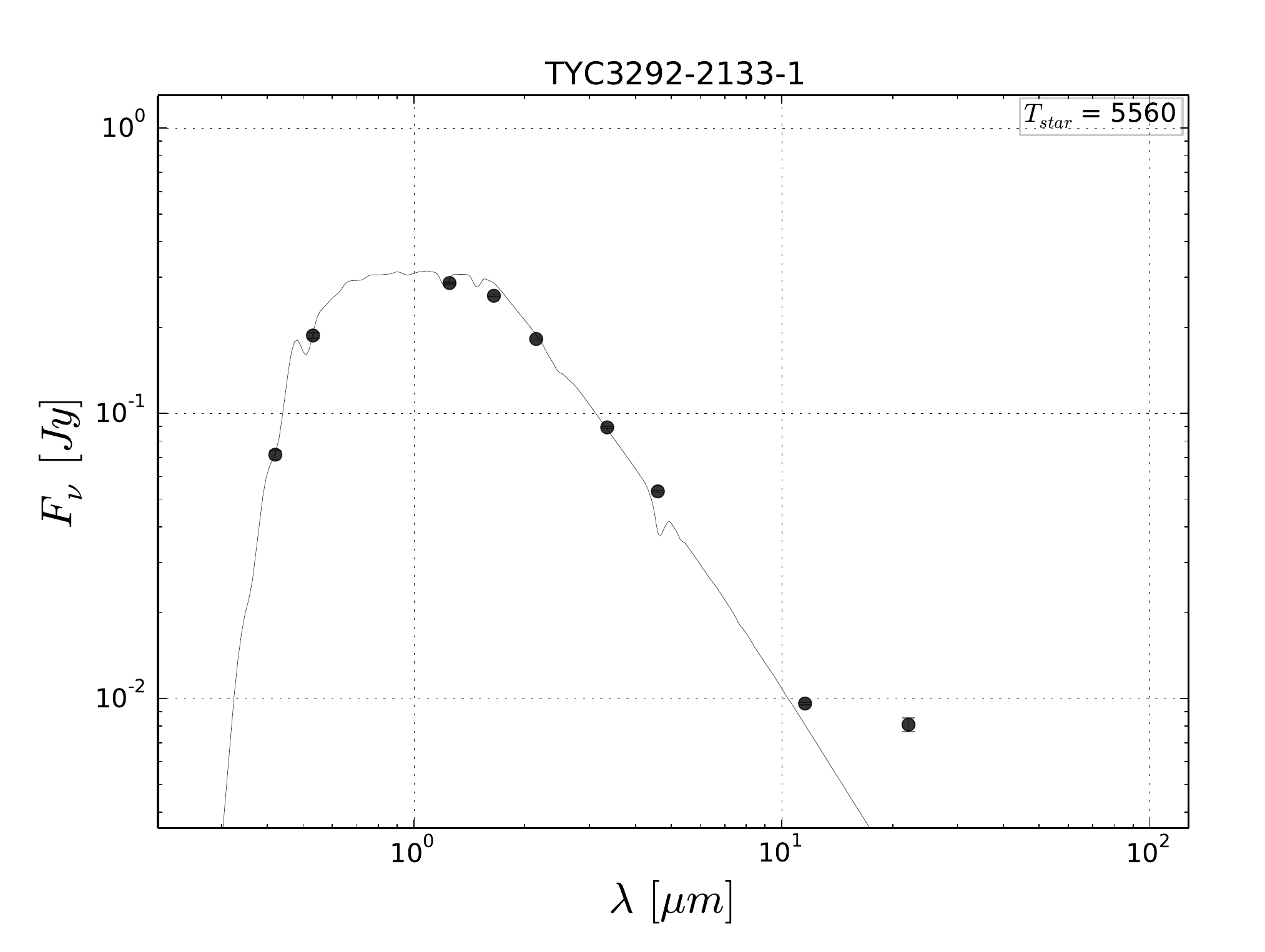} & \includegraphics[height=2.2in]{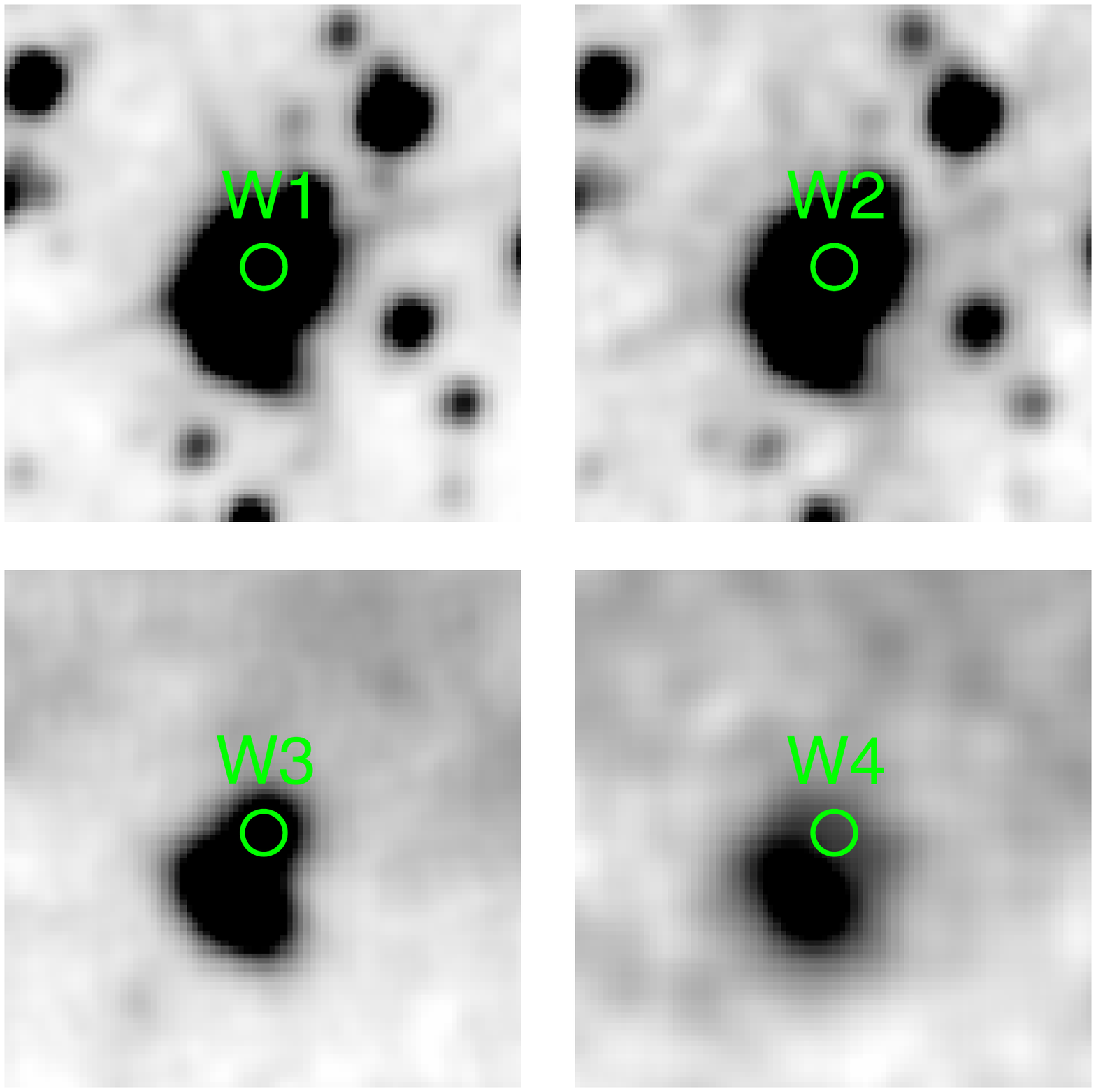} \tabularnewline
b. & \includegraphics[height=2.2in]{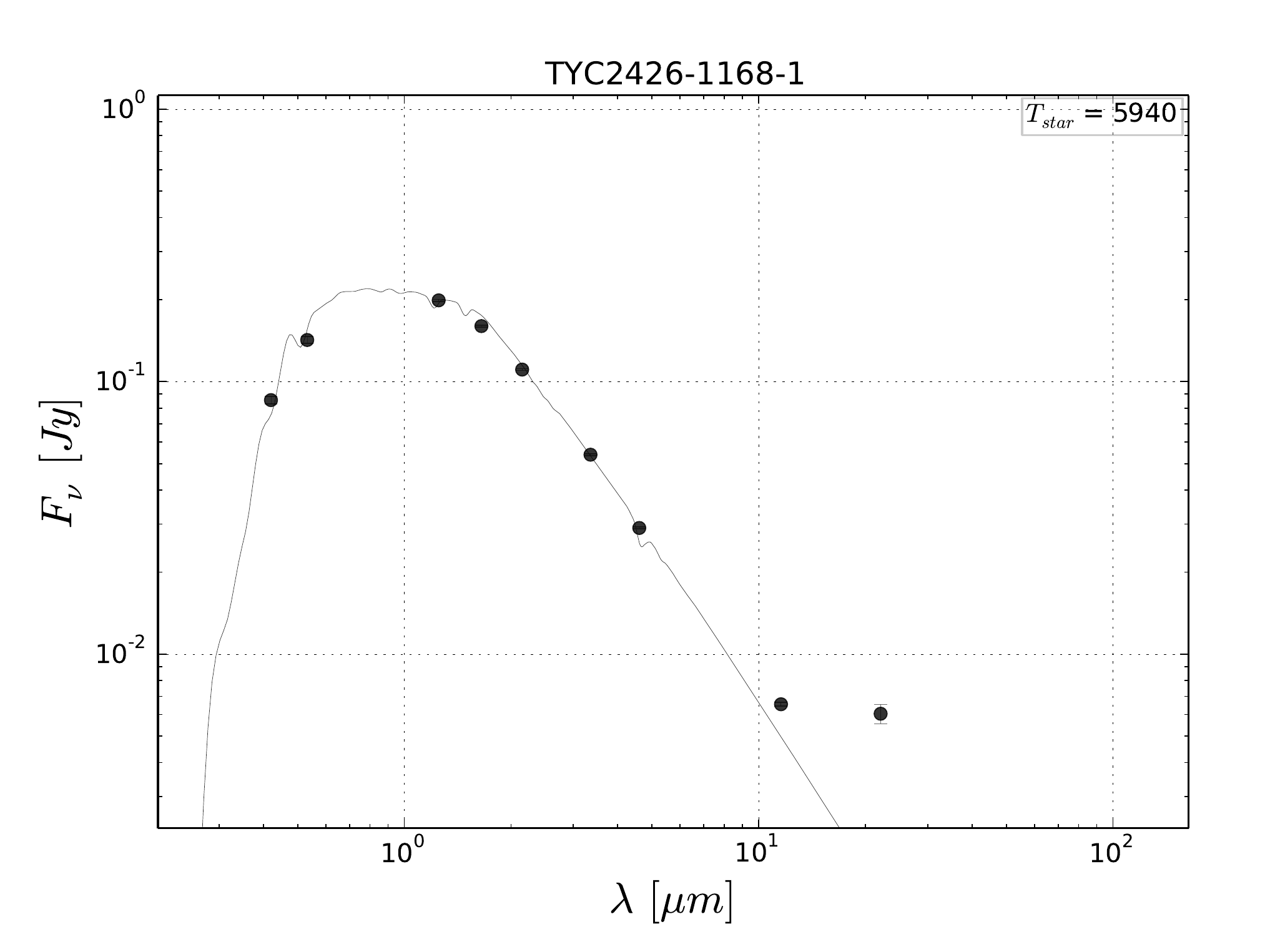} & \includegraphics[height=2.2in]{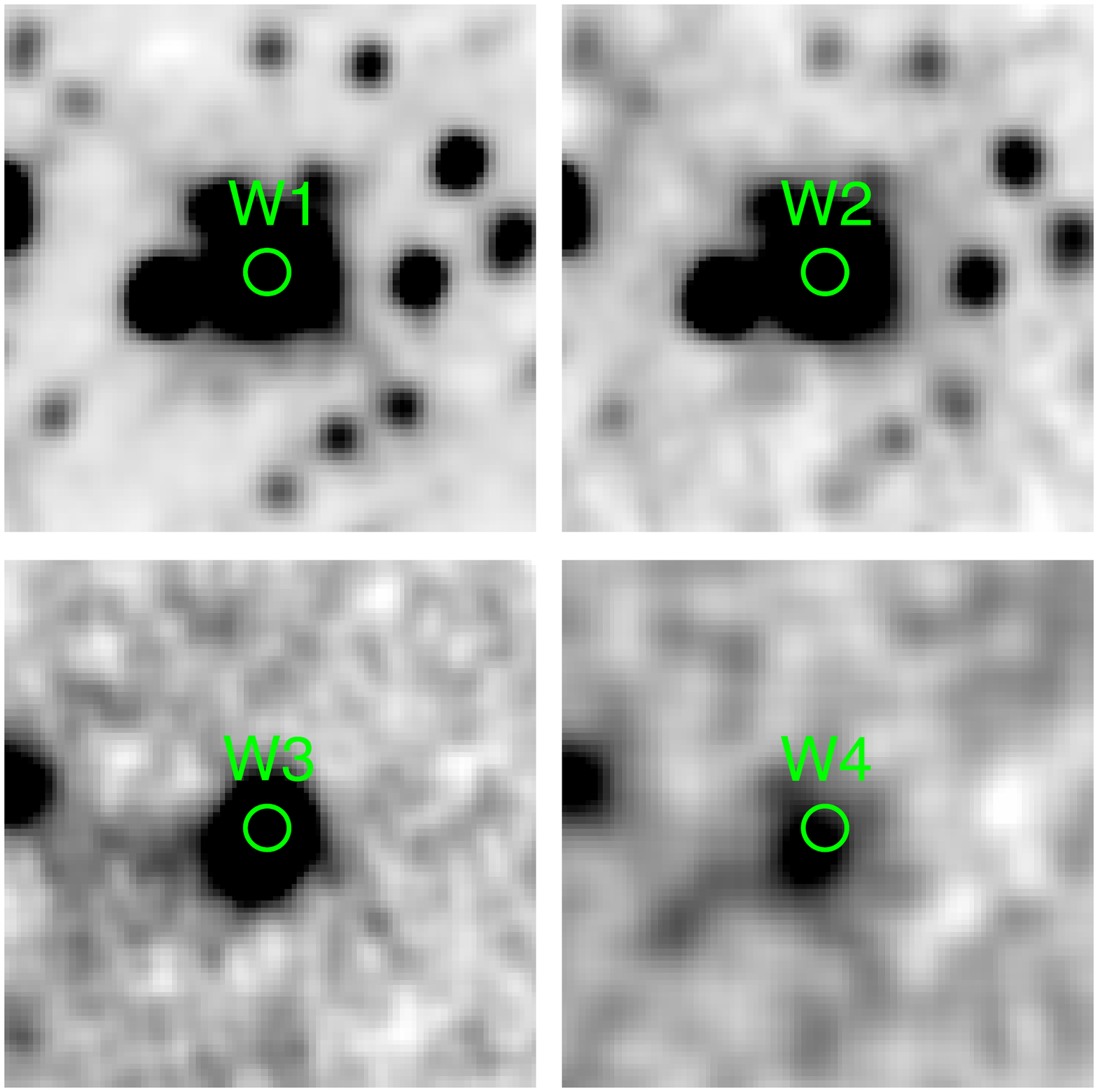} \tabularnewline 
c. & \includegraphics[height=2.2in]{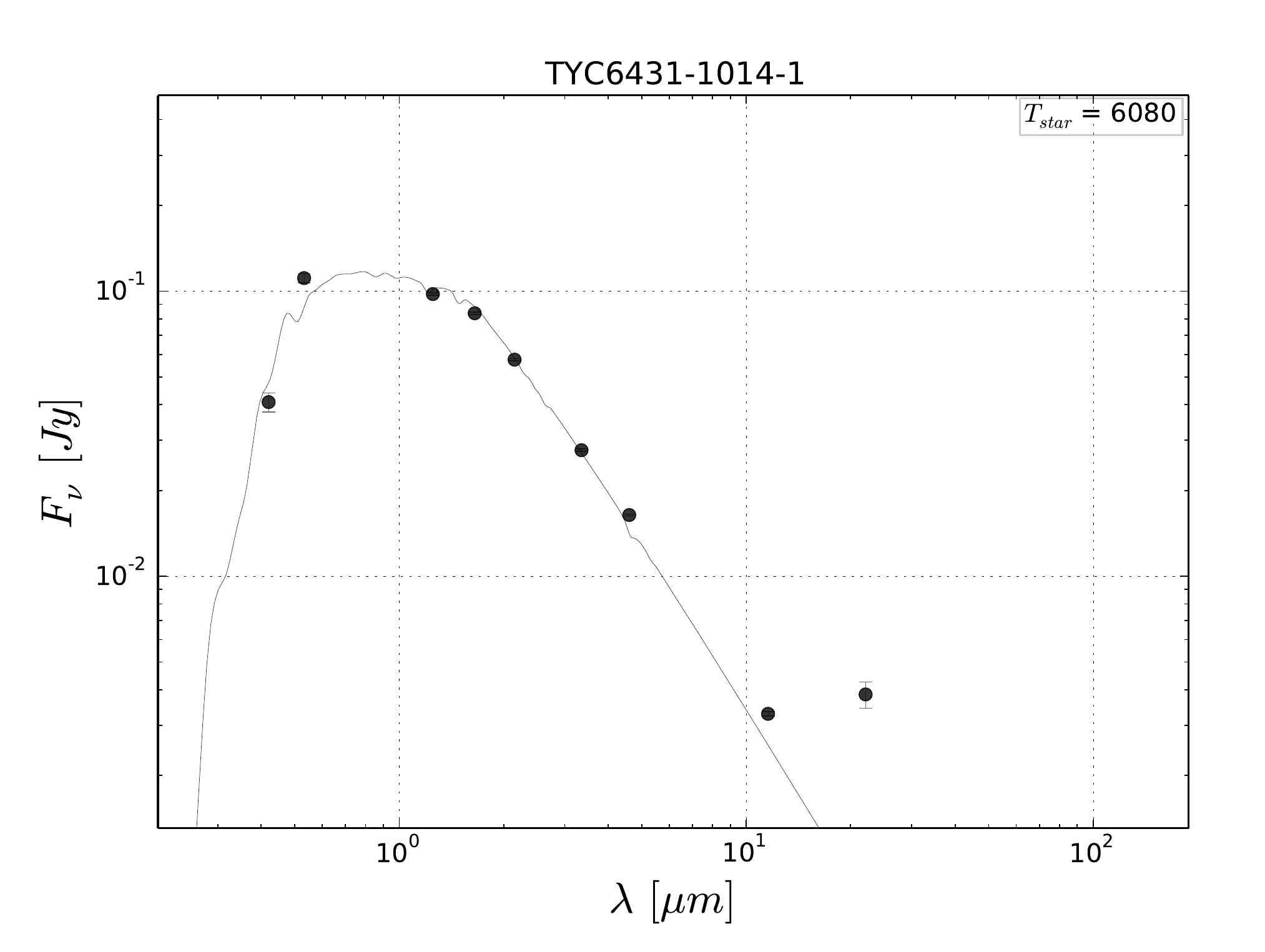} & \includegraphics[height=2.2in]{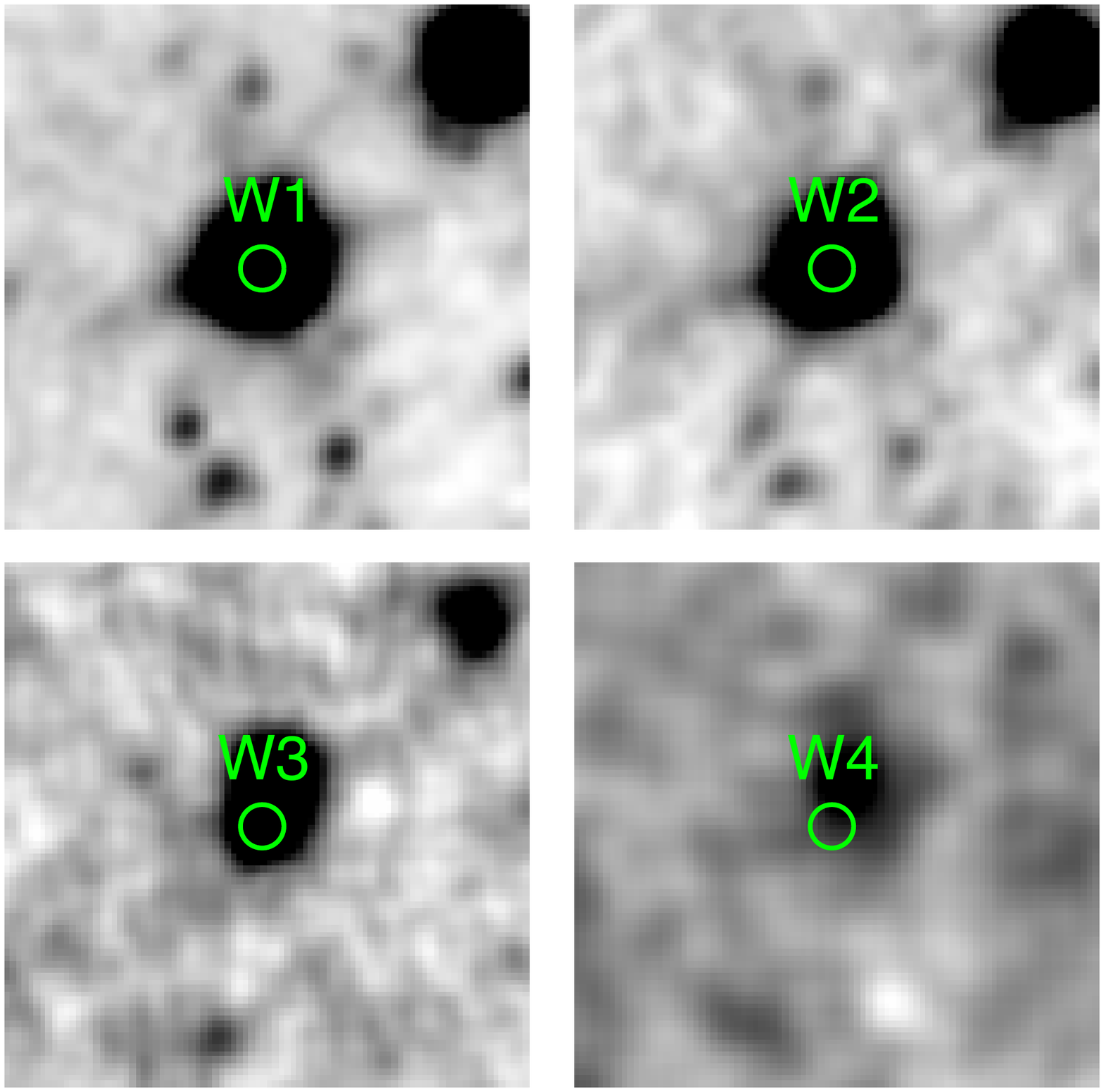} \tabularnewline
\end{tabular}
\protect\caption{This figure shows examples of stars removed
during the image vetting procedures. The \emph{left-hand} figure displays the SED for each star with the photosphere fit to Tycho-2 B$_{T}$, V$_{T}$, 2MASS J, H and K$_{S}$. 
The \emph{right-hand} figure shows the AllWISE images taken from the NASA/IPAC Infrared Science Archive and are 2\arcmin{} by 2\arcmin{}, scaled linearly. 
The images also contain a 5.0\arcsec{} circle centered on the search position. 
a.$)$ Due to both the background
contaminating cirrus and the large offset to the brightest source
in W4, this source is likely a background galaxy found through
our offset criteria.
b.$)$ This source passed the initial offset criteria,
but further inspection proved sources with W4 offsets $>$ 8.0\arcsec{} needs
to be removed as well. The object shown in the W4 image is offset from
the W2 position by 9.1\arcsec{} and yet the amount of W4 excess emission 
cannot be due to a center source alone. 
c.$)$ This source demonstrates an elliptical shape in W3, this
source is rejected based on the roundness criteria and is likely
a background IR source. 
See Section \ref{visualcut} for more details regarding these images.}
\label{removedex}
\end{figure}
\end{center}

To implement the image inspection, we download 2\arcmin{}$\times$ 2\arcmin{} images at each AllWISE
passband followed by the source detections on these images using IRAF's program, \emph{daofind}. We compare 
detected source positions and shapes against the expected stellar positions.
All sources were detected in W3, but 770 targets went undetected in W4. Of these non-detections, 92$\%$ are
cirrus contamination leading to spurious excess fluxes, while the
remaining are extended sources indicating large nearby
objects, most likely galaxies. The remaining sample contains 3530 sources with identified
\emph{daofind }positions in each AllWISE passband that are plotted in Figure \ref{posoffset}
displaying the detected offset between the source position in W3 or
W4 compared to W2 since W2 most accurately reflects the Tycho-2 source position. 
First, we remove sources with positional offsets greater than the resolution of W3 (6.7\arcsec{}) or W4 (12.0\arcsec{}) in AllWISE.
Upon further inspection, however, sources offset in W4 greater than 8.0\arcsec{} are also contaminated. 
Removing 120 sources from offsets in W3 preserves 95\% of the sample, but separately, 
1180 targets are removed for large offset positions in W4. In this
case, 44$\%$ of the sources appear to be cirrus, 25$\%$ are due
to background sources, and the remaining 30$\%$ are sources
unresolved in the AllWISE images in which the secondary object is brighter
at this passband and so shifts the detected source position. 
Further, excess candidates were inspected for potential ellipticity.
Of the 150 sources we remove, 64$\%$ appear to be cirrus, 22$\%$
are some sort of background object or nebulae, and 14$\%$ are likely double
stars of similar brightness that are unresolved at W3 or W4. The double
stars are removed here because the secondary source will contribute
additional flux to the target star and act as a false-positive IR excess
candidate. The candidate IR excess stars now total $\sim$2100. 
Examples of some contaminated targets are displayed in Figure \ref{removedex}. 

We mentioned previously, we expect $\sim6\%$ of the remaining candidates to
be giants based on our discussion in Section ~\ref{giants}.
With the sample now reduced, we search each candidate position in
SIMBAD to identify a published luminosity class or a contaminating object
nearby such as a background galaxy. 
We remove 81 stars with luminosity classes of I, II, or III. Through this endeavor, we also
remove 46 other contentious objects such as Cepheid variables, white
dwarfs, nebulae, novae, and known galaxies (within 10\arcsec{}). Additionally, we eliminate
523 sources through inspection of the 2MASS images which display
are likely double or multiple star systems (within 10\arcsec{}) unconfirmed in the 
AllWISE images, especially W4. 
Most double stars that are unresolved in the WISE passbands show excess above the photosphere from the over-estimated flux in each
passband, and yet still mimics a Rayleigh-Jeans tail.
Through inspection of the SED of each star for confirmation or rejection of likely binary-related false excess.
The IR excess sample reduces to 1430 stars.

In order to affirm the SED fit and IR excess, we also gather additional
photometry through VizieR from the ultraviolet to the far-IR (including
data from \emph{Spitzer }MIPS and the \emph{Herschel}
PACS or SPIRE instruments; \citealt{poglitsch2010}; \citealt{griffin2010}) 
and performed a re-evaluation of the candidate SEDs. Additional photometry did one of three
things in terms of contamination or true source identification: 1$)$ it shifts
the stellar photosphere slightly such that a star's significance of excess, now no longer
passes our criteria, 2$)$ it provides additional far-IR photospheric data which refutes the AllWISE excess, 
or 3$)$ it corroborates previous excess detections.
Far-IR data from the \emph{Spitzer }MIPS
instrument provides better sensitivity through pointed observations, therefore, we treat the \emph{Spitzer} 
data as representative of the true measurement of far-IR flux.
The third case of additional photometry is the only example of beneficial photometry, and so we will 
examine the other cases in more detail.
Firstly, we include more data from the optical and near-IR region of the SEDs from VizieR. 
More data improves the fit to the stellar photosphere 
for $\sim$80 stars, which shifts the photosphere and the infrared Rayleigh-Jeans tail and a reevaluation of the 
significance of excess makes these sources non-excess.
Secondly, additional photometric measurements at far-IR wavelengths (MIPS, PACS, SPIRE) can reject marginal 
cases of mid-IR excess.  We removed 60 objects with 
photospheric detections at MIPS 24 and no IR excess at 70$\mu$m. 
However, one interesting case, HD 69830, found by \citet{beichman2005b} has mid-IR excess around 10$\mu$m confirmed 
using interferometric evidence by \citet{smith2009b} without the presence of any far-IR excess detection.  
This rare and unique circumstellar disk is also host to three exoplanets \citep{lovis2006}.
While we do not want to miss any interesting cases, we removed 60 additional stars that have the potential to be 
HD 69830-like due to true mid-IR excess in W3 and/or W4 and far-IR photospheric detections, 
since we believe the majority of which are likely bogus excess sources.
Our IR excess candidate sample comprises 1230 sources.

\subsubsection{Reliability of Excess and Final Distance Restriction}\label{distcut}

Proximity to our Sun serves as a initial quantifier for our `Prime' targets.  
We initially selected stars using the total proper motion magnitude as
a proxy for a distance corresponding to $\sim$200 pc. Considering the size of the Local
Bubble and the minor ISM influence on the IR excess in this region,
we place stars within 120 pc in our Prime catalog.  
This identified $\sim$300 stars from our Tycho-2/AllWISE IR excess sample with measured distances. 

For the remaining 67\% of stars without trigonometric parallaxes,
we develop a more reliable distance restriction based on a photometric
distance calculated by our SED fitting algorithm. 
The SED fit can use all photospheric photometry for which the brightnesses are highly covariant instead 
of a single photometric passband.
This procedure is based
on the final step of the SED fitting algorithm which multiplies 
the model fluxes by a scale factor, namely $(\frac{R}{d})^{2}$,
where $R$ refers to the stellar radius in solar radii 
and $d$ is the distance to the star in parsecs (see discussion in \citealt{cushing2008} for
more details). 
We then investigate deducing a distance from this scale factor value using model isochrones
to infer an expected stellar radii given the best fit stellar temperature
returned from the SED fit. Using the isochrones from \citet{siess2000},
\citet{allard2011}, and \citet{pisamodels2012}, we construct a
composite isochrone model which ranges between 2000 and 15000K (for more details
regarding this model, see Lee \& Song 2016, in prep.). 
SED distances are estimated for about 900 stars and the expected uncertainties is typically $\sim12\%$. 

The effort to include the SED distance for stars without trigonometric
parallax identified 48 stars that are likely evolved M-type stars. 
Inspection of their SED distance within 15 pc and the total
proper motion (less than 40 mas/yr) offers conflicting data regarding the location of these stars.
Although there are two small regions (solar apex and anti-apex) where nearby stars 
exhibit very small proper motions, 
if the total proper motion is $\sim$40
mas/yr, we expect the distance to be on the order of 100 pc. 
However, for these 48 stars in our sample
the SED distances of less than 15 pc are very unrealistic. 
We retain these stars in the Reserved sample with an accompanying note until spectroscopy can
confirm that they have indeed evolved away from the main sequence.

Finally, to aid in the reliability of our IR excess stellar sample,
we generate a parameter called `Num\_Excess' in Tables \ref{highfidelity}
and \ref{reserved}, that represents the number of passbands
which display excess above the photosphere.
We designate the following wavelength ranges as our IR excess passbands: 5 - 13$\mu$m, 
17 - 30$\mu$m, 55 - 75$\mu$m, 90 - 110$\mu$m, 120 - 170$\mu$m, 200 - 300$\mu$m, 
300 - 400$\mu$m, 400 - 500$\mu$m, 500 - 600$\mu$m, 700 - 900$\mu$m, and $>$1000$\mu$m. 
Thus, instruments with similar wavelength photometry such as WISE W4 (22$\mu$m), 
\emph{Spitzer} MIPS 24$\mu$m, and IRAS 25$\mu$m are maintained as corroborative evidence of the same excess near 25$\mu$m.
Besides being more reliable indication of excess, multiple excess detections reduce 
the inherent degeneracy in fitting a blackbody to the dust temperature, thereby allowing for a more detailed dust analysis. 
Tables \ref{highfidelity} and \ref{reserved} will display Num\_Excess, the starting wavelength of the IR excess in microns,  
and a flag if the \emph{Spitzer} IRS spectra is available. 
These columns should provide a full description of the nature of the excess.
We also include in the online materials a table of photometric 
measurements and model predictions at key stellar fitting passbands for all the stars in Tables 
\ref{highfidelity} and \ref{reserved} as well as the measured photometry from \emph{Spitzer} and/or 
\emph{Herschel} if available from the literature. An example of the content and form of the photometry table is provided here 
in Table \ref{phottable}.

We compare our final tables of Tycho-2/AllWISE Prime IR excess stars and Reserved IR excess candidates 
with the literature sample found in Section \ref{litsearch} in order to exclude duplicate entries. 
Then, the following reliability criteria are used as in Section \ref{litsearch} to select reliable 
IR sources from the Tycho-2/AllWISE sample for our Prime catalog: 
either
\begin{enumerate}
\item at least 2 passbands demonstrating IR excess (`Num\_Excess'$>$1), AllWISE W3 or W4 flux greater than 10 mJy
(for more efficient follow-up observations), and proximity such that distance is within 120 pc; 
\item at least 1 passband demonstrating IR excess, AllWISE W3 or W4 flux greater than 10 mJy, 
has IRS spectra confirming photometric IR excess identification, and distance $<$120 pc.
\end{enumerate}
Passing the above criteria, we compile just about 500 stars from the Tycho-2/AllWISE search and the literature 
search for the Prime IR excess catalog. 
The Reserved table contains over 1200 stars\footnote{The full extent of Tables \ref{highfidelity} and 
\ref{reserved} are available as online material for this manuscript as well 
as stored locally on a public server through the University of Georgia entitled: 
Debris Disk Database (www.debrisdisks.org).}.

\section{Previous WISE Excess Searches}\label{comparison}

This study is not the first use of the WISE survey to search for IR excess,
therefore we compare our results to the debris disk candidates
discovered by \citet{wu2013}, \citet{cruzsaenz2014}, \citet{patel2014}, \citet{theissen2014} and \citet{vican2014}. 
We will avoid a comparison between our final samples and \citet{rizzuto2012} since they 
made use of the preliminary WISE catalog data only which is now obsolete once the full catalog was released 
and many of their stars were reassessed through other studies.
Each of these searches maintains a similar sample criteria to our new Tycho-2/AllWISE
cross-correlation, however, each study presents a different standard
of significance of excess and final analysis of the candidates.

First, \citet{wu2013} was specifically searching for 22 $\mu$m excess
using \emph{Hipparcos} stars within 200 pc. 
They selected candidate excess stars using a color criteria of {[}K$_{S}-$W4{]}
and produced a sample of 141 candidate IR excess stars. 
Our study was able to reproduce 63 stars
from the final \citet{wu2013} IR excess sample. From the remaining
stars that were not matched, 44\% have total proper motions according
to the Tycho-2 catalog that are less than 25 $mas/yr$ (the initial
sample selection that we used) and the rest (56\%) do not pass our
significance of excess. 

\citet{patel2014} found 220 stars showing W3 and/or W4 excess. 
We confirmed 114 ($\sim52\%$) of their IR excess stars. 
Majority (85\%) of the remaining 106 IR excess candidates do not pass our significance
of excess criteria and the rest (15\%) show total proper motions less than
25 $mas/yr$ implemented as the initial search criteria of our study.

Constraining their sample using SIMBAD, \citet{cruzsaenz2014} selected only
main-sequence dwarfs and any distance. 
They use a comparison between the predicted and measured ratio of fluxes (W4/W2) divided by the calibration weighted uncertainty.  
Through the typical (and somewhat subjective) process of visual inspection of WISE images
to remove false-positive IR excess candidates, \citet{cruzsaenz2014}
report 197 IR excess stars. We confirmed 40 of these sources
as IR excess stars in our Prime or Reserved catalogs. 60\%
of the \citet{cruzsaenz2014} sample do not pass our significance
of excess cut and the remainder (40\%) do
not pass our total proper motion criteria. 

Previous searches for M stars with IR excess by \citet{avenhaus2012}, who used the AllWISE catalog and 
the RECONS sample from \citet{henry2006}, did not find any new cases of M stars with IR excess indicating a debris disk. 
\citet{theissen2014} were able to expand this search using the Sloan Digital Sky Survey (SDSS) 
spectroscopically selected M-type stars out to a distance of nearly 2000 pc.  While they 
executed a series of contamination checks to ensure their sample contains only M dwarfs, only 36\% of their 
175 IR excess stars are within 200 pc. None of these stars were reproduced by our search.  The main reason 
for the dissimilar results with respect to our sample is distance. Moreover, of the 63 M dwarfs within 200 pc,  
none of them have Tycho-2 catalog matches.  The faintness limit of Tycho-2 would exclude many M dwarfs at these 
distances. 

\citet{vican2014} searched the literature for stars with chromospheric
activity indicators in order to constrain the age of their sample before using 
WISE to determine if these stars had IR excess.  Their criteria for excess 
is very similar to our significance of excess using the comparison between measured 
and photospheric fluxes, and they found  
98 IR excess candidates. 
Twenty-four of these sources can be found in our Tables \ref{highfidelity} and \ref{reserved}. 
Eight stars were not selected to our sample due to very low total proper motion and the rest (66)
were not selected based on our significance of excess criteria.  

\section{Sample Characteristics}\label{samplechar}

The Prime IR excess star catalog (Table \ref{highfidelity}) contains $\sim$500 
stars and information regarding the best fit SED stellar
effective temperature, the number of excess passbands, distance, and
disk parameters such as temperature, radius, and fractional dust luminosity.
We have marked the previously published IR excess stars using the
column ``known'' that designates one of the star's first claim of IR excess
in the literature. Table \ref{reserved} mimics the display of the Prime catalog.
Most notably, this study has produced $>$70 new Prime IR excess stars which is almost 
a 20\% increase in notable IR excess stars as well as a 
handful of very dusty disks ($L_{IR}/L_{*}\equiv\tau>10^{-2}$).
Further, the number of marginal IR excess stars has quadrupled with this new Tycho-2/AllWISE search.

For the disk parameters, we choose the most simplistic model
to define the disk temperature and radius. We
assume the SED can first be fit with a single blackbody function.
If the single blackbody model does not fit the dust, then a two-blackbody disk
model is applied. Stars with \emph{Spitzer} IRS spectroscopy have disk models that match the spectroscopic evidence.  
In addition, for stars in the Reserved catalog with IR excess in only one passband and without \emph{Spitzer} 
IRS spectroscopy, we assume a warm dust disk whose dust flux peaks at the single excess wavelength.
The fitting parameters are listed in Tables \ref{highfidelity} and \ref{reserved}. 

Past studies presented several estimates on the occurrence rate of excess stars focusing
on nearby, main sequence stars and in some cases focusing on a specific
cluster, region, or spectral type. 
\citet{su2006} report a debris disk fraction of $\geq$ 33$\pm$ 5$\%$ surrounding A stars using \emph{Spitzer }MIPS measurements of excess
emission at 24 or 70 $\mu$m while 
\citet{beichman2006b} combined their results with the \citet{bryden2006a} and \citet{gautier2007}
studies to report 15$\pm$ 3$\%$ of F0 - K0 type stars have debris
disks detected at 70 $\mu$m. 
\citet{gorlova2006} surveyed the 30 Myr old cluster NGC 2547 and report incidences of 40$\%$ of B to F type stars demonstrate
excess at 24 $\mu$m. 
More recently, \citet{eiroa2013} discussed the overall prevalence of far-IR 
detection of circumstellar dust around nearby solar-type stars to be nearly 23\%.
This can be contrasted with larger all-sky searches
for nearby solar-type stars with mid-IR excess such as those performed by \citet{wu2013} and \citet{patel2014}
who found incidence rates of mid-IR excess specifically at 22$\mu$m 
solar-type stars of 2.21$\%$ and 1.8$\%$, respectively, 
To create a comparable sample of stars which are 
not contaminated by nearby objects, we filter the 250,000 starting sample of Tycho-2/AllWISE stars to exclude giants (originally $\sim$30\%) 
and then eliminate the expected number of sources with nearby contamination ($\sim$50\% of 200,000 stars) 
extrapolated from the results in Section \ref{visualcut}.
This procedure interprets $\sim$120,000 main sequence stars from various sources 
that have the potential to contain IR excess stars without any contaminating influences. 
From this sample, we determine the occurrence of IR excess for 
Tycho-2/AllWISE stars with W4 excess from the parent sample is 1.0\% (1200/120000).
In order to best compare with past IR excess studies, we focus on the incidence of 22 $\mu$m (W4) excess 
around A-type or earlier stars and solar-type stars only.
Excluding the literature sample from this analysis, we find that 15.6\% ($\sim$320 of 2100 stars 
with SED temperatures that reflect A-type from the 120000 sample) have infrared excess at W4, while 
the solar-type stars have an incidence rate for W4 excess of 0.7\% ($\sim$850/115000).
This value is in agreement with the values put forth by \citet{wu2013} and \citet{patel2014}, but still less.
We will avoid a discussion of M-type IR excess incidence until spectroscopic evidence indicates a main sequence age. 

\begin{center}
\begin{figure}
\begin{centering}
\includegraphics[width=6.5in]{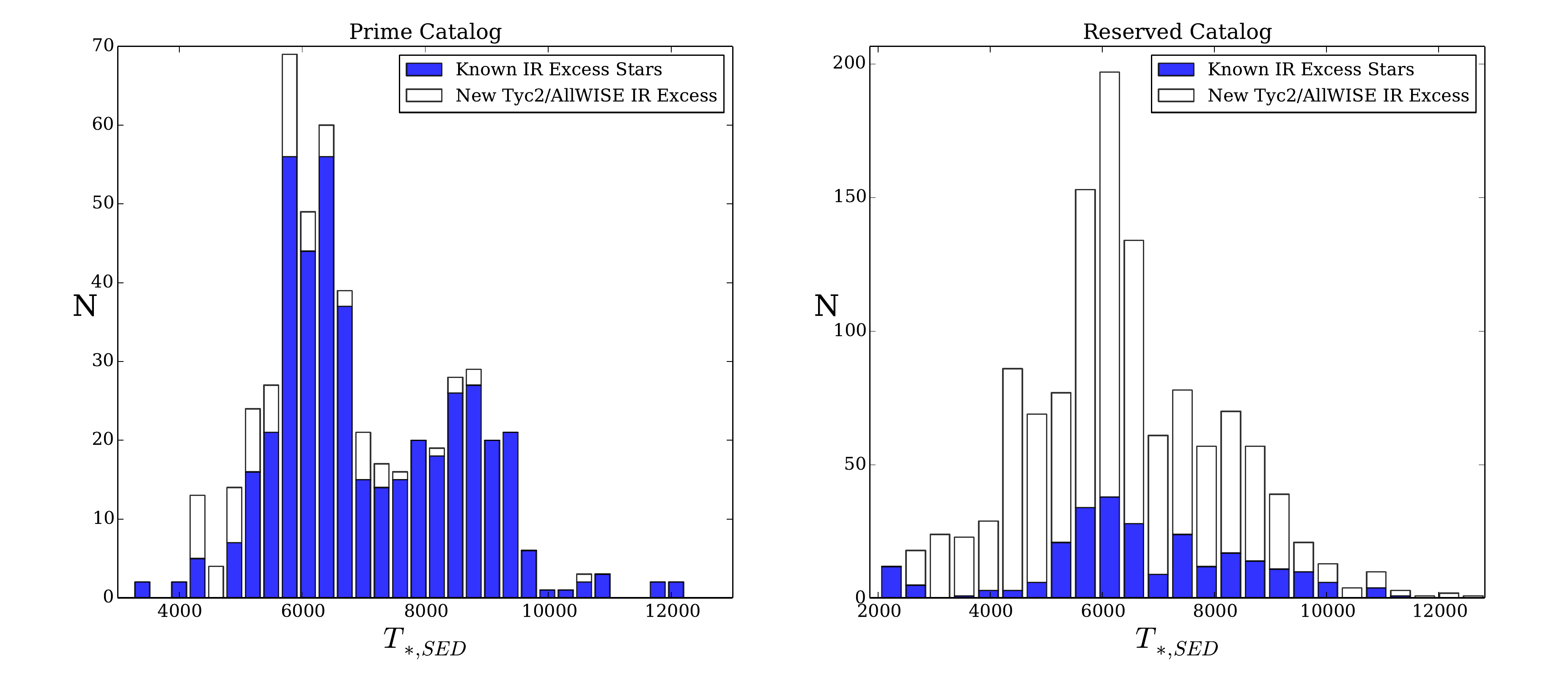}
\caption{Distribution of best fit stellar temperature in
the final, Prime IR excess sample (\emph{Left}) and the Reserved sample (\emph{Right}). 
The filled histogram contains
stars previously claimed to display IR excess and reproduced in this
study. The unshaded histogram contains stars which are new IR excess detections
from our Tycho-2/AllWISE search. Refer to Section ~\ref{samplechar}
for more details. }
\label{sptypehist}
\end{centering}
\end{figure}
\end{center}

Considering the spectral types of the Prime and Reserved catalogs, Figure \ref{sptypehist} shows the range 
of the best fit SED stellar temperature. 
F and G type stars show the largest prevalence of IR excess in these catalogs.
Given the sensitivity to detect infrared excess depends on spectral type and distance, the distribution of 
Prime IR excess stars, especially the apparent concentration of F and G-type stars, is likely an observational bias. 
However, F and G stars are important for understanding the formation and evolution stellar systems analogous to our Solar System. 
As a secondary check for our use of the SED effective temperature as a spectral type, we 
gather spectral type information from SIMBAD where the luminosity class has already been determined (82\% of the Prime IR excess catalog). 
We use the relationship between the spectral type and the SED stellar temperature in order to 
extrapolate a spectral type for the remaining stars.  
Without significant outliers, we conclude that our SED fitting algorithm is working effectively.  
Both Tables \ref{highfidelity} and \ref{reserved} will display the spectral type 
either with a luminosity class from SIMBAD or with a `:' signifying it is an interpolated spectral type from SED fits. 

\begin{center}
\begin{figure}
\begin{centering}
\includegraphics[width=6.5in]{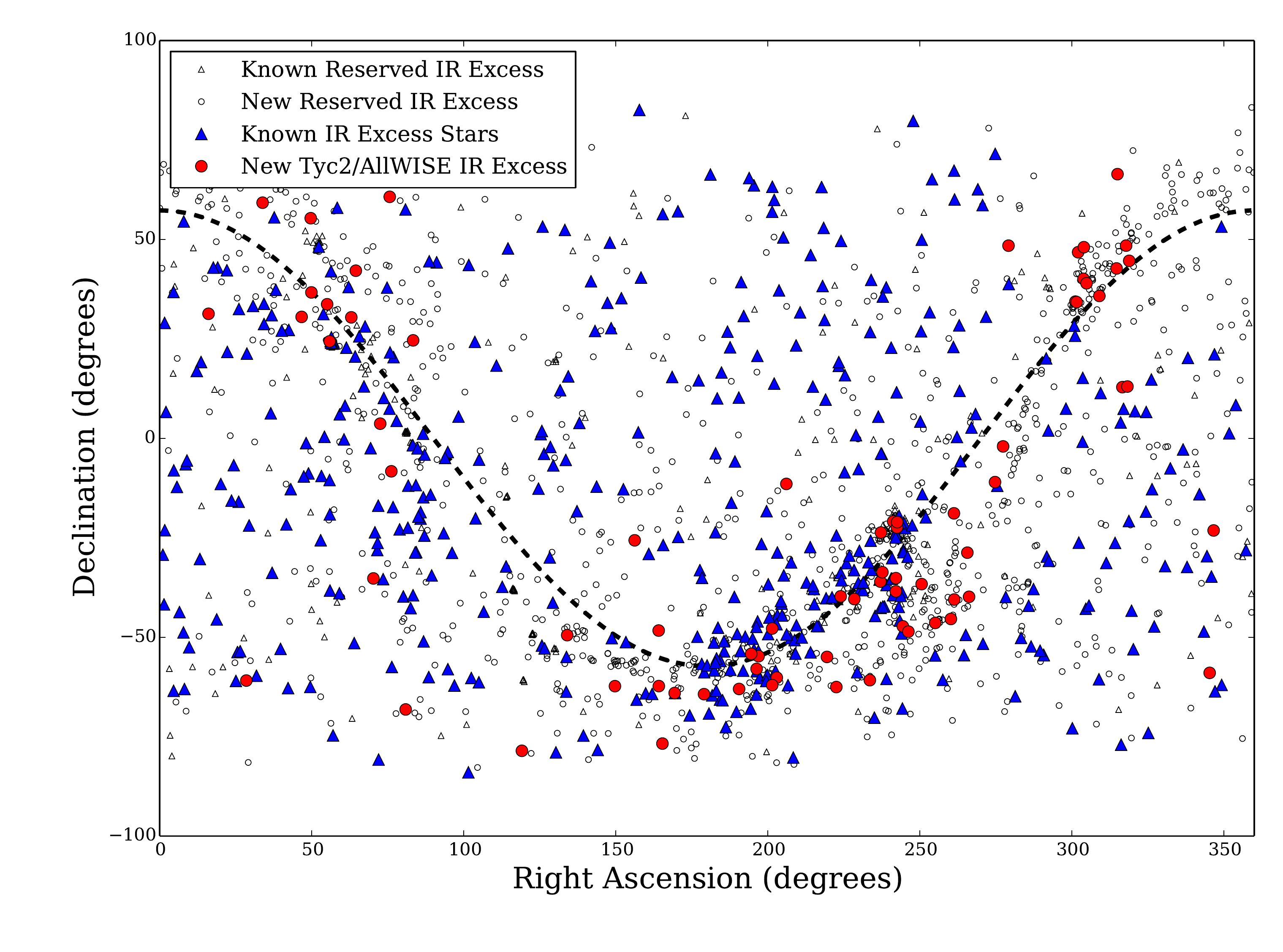}
\protect\caption{Spatial distribution of IR excess stars in R.A. and Dec.
Red circles represent the new, Prime Tycho-2/AllWISE IR excess 
stars and blue triangles show the sample of known, Prime
IR excess stars. The grey symbols are from the Reserved catalog that did not
qualify for the Prime targets. The symbol shapes are the same for
the literature and new IR excess stars in the Reserved sample.
For reference, the dashed curve signifies the galactic plane.}
\label{radec}
\end{centering}
\end{figure}
\end{center}

We confirm in Figure \ref{radec} the representation 
of IR excess stars across the entire sky. 
The dashed curve indicates the galactic plane. 
Stars centered on the galactic plane may show IR excess as the result 
of source confusion and likely introduce false positives into the sample.
However, since the new IR excess stars are not crowded around the galactic plane, 
we reason that the source confusion did not jeopardize our final catalogs. 
The two different symbols represent the literature excess
sample (triangles) and the new IR excess stars (circles) found here. The small unfilled symbols
represent the entire Reserved catalog from Table \ref{reserved}.
We note in Figure \ref{radec} that there are more candidate
IR excess stars at declinations less than -10$^{\circ}$ and at right
ascensions between 150$^{\circ}$ and 250$^{\circ}$. This region
highlights the Scorpius-Centaurus Association (ScoCen, hereafter). Based on the age of
$\sim11$ Myr approximated by \citet{pecaut2012}, we expect this
region to have an enhanced density of younger stars with
circumstellar material, however, the ages of the ScoCen subregions are still 
not fully settled (see \citealt{herczeg2015}). The new IR excess stars that are located near the ScoCen region
may be on the forefront of this massive population of young stars.

\section{Discussion}\label{discuss}

We have begun a detailed characterization study 
of the Prime stars by gathering optical spectroscopy 
and will be described in a future manuscript studying the
relationship between stellar parameters and the dust. Here, we examine
the dust parameters in order to investigate
relationships between the SED fitting parameters: stellar temperature,
dust temperature, dust radius, and the fractional dust luminosity.
Lastly, we discuss the relationship between the SED disk 
radius and the true size of the disk for those disks resolved with scattered light imaging.

\subsection{Fractional Dust Luminosity}\label{lir_lstar}

\begin{center}
\begin{figure}
\begin{centering}
\includegraphics[width=6.5in]{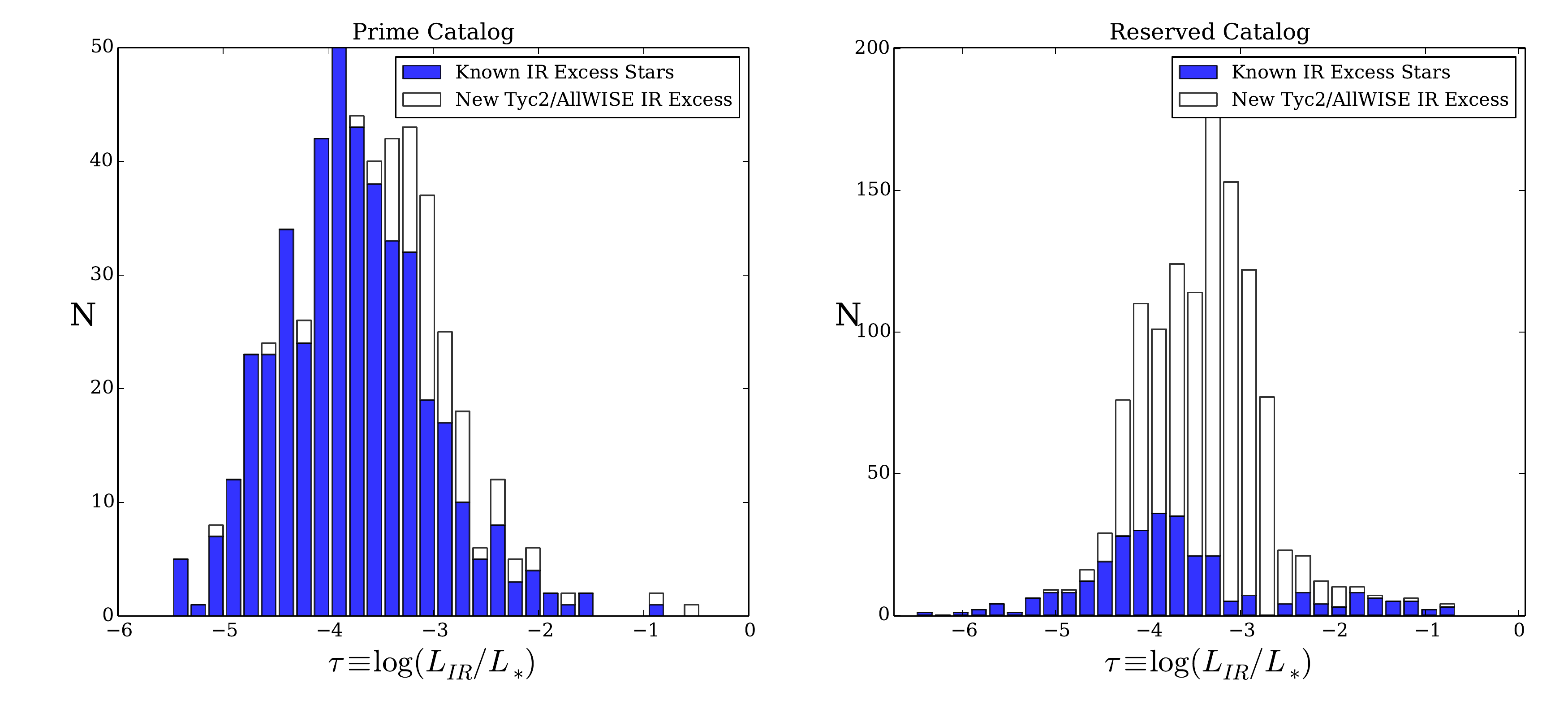}
\protect\caption{Histogram of the fractional dust luminosity displayed in logarithmic scale for the Prime IR excess 
catalog (\emph{left}) and the Reserved catalog (\emph{right}).  The average
fractional dust luminosity for the Prime catalog has a value of $10^{-3.8}$ where low $\tau$ stars were not detected due to 
limited sensitivities of IR excess surveys.  There are a handful of extremely dusty disks
($\tau > 10^{-2}$) which will be interesting for further understanding the formation and evolution
of dust around a star.}
\label{tauhist}
\end{centering}
\end{figure}
\end{center}

The fractional dust luminosity defined as $\tau\equiv L_{IR}/L_{star}$,
provides an estimate for the amount of dust surrounding each of the
IR excess stars regardless of the number of blackbody fits to the dust. 
Figure \ref{tauhist} shows the distribution of fractional dust luminosity for the Prime
catalog (left) and the Reserved catalog (right). 
From this distribution, we see that the majority of our main sequence
IR excess stars have $\tau$ values between $10^{-5}$ and
$10^{-3}$. The Reserved sample shows a more central peak around $10^{-3}$, although 
this peak should be evaluated with caution since many of the reserved IR excess stars 
only display one passband of excess which increases the likelihood of degeneracy in the 
SED fit to the dust and may select dust temperatures which suggest a higher fractional dust luminosity.
Reflecting on our sample, \citet{roberge2012} explored the sensitivity limits of the fractional dust luminosity of past instruments. 
However, this investigation of the limiting detectable fractional dust luminosity assumes that each IR excess 
star has the same amount of dust and is representative of IR excess at far-IR wavelengths. 
The detectability at mid-IR wavelengths is complicated by the necessity for precise photospheric 
detections from which to distinguish IR excess and so the limit of mid-IR surveys is more challenging to assess.
While not found in our sample and outside of the currently detectable range (i.e. $\tau<10^{-7}$), 
we would expect extremely dust poor debris disk systems 
similar to our solar system ($\tau\sim10^{-7}$; \citealt{wyatt2008}, \citealt{moromartin2008}, \citealt{vitense2012}, \citealt{nesvorny2010}).
Yet, the other extreme in Figure \ref{tauhist} (i.e. $\tau>10^{-3}$) provides interesting objects since the amount of dust can either be the result
of a large, destructive, transient event such as a period of Late Heavy Bombardment (LHB) 
or is sustained by the support generated due to presence of gas.  
A period such as LHB would indicate a mature system that we are able to observe at a crucial planet evolution epoch
while the gas may indicate extreme youth. 

\begin{center}
\begin{figure}
\begin{centering}
\includegraphics[width=6.5in]{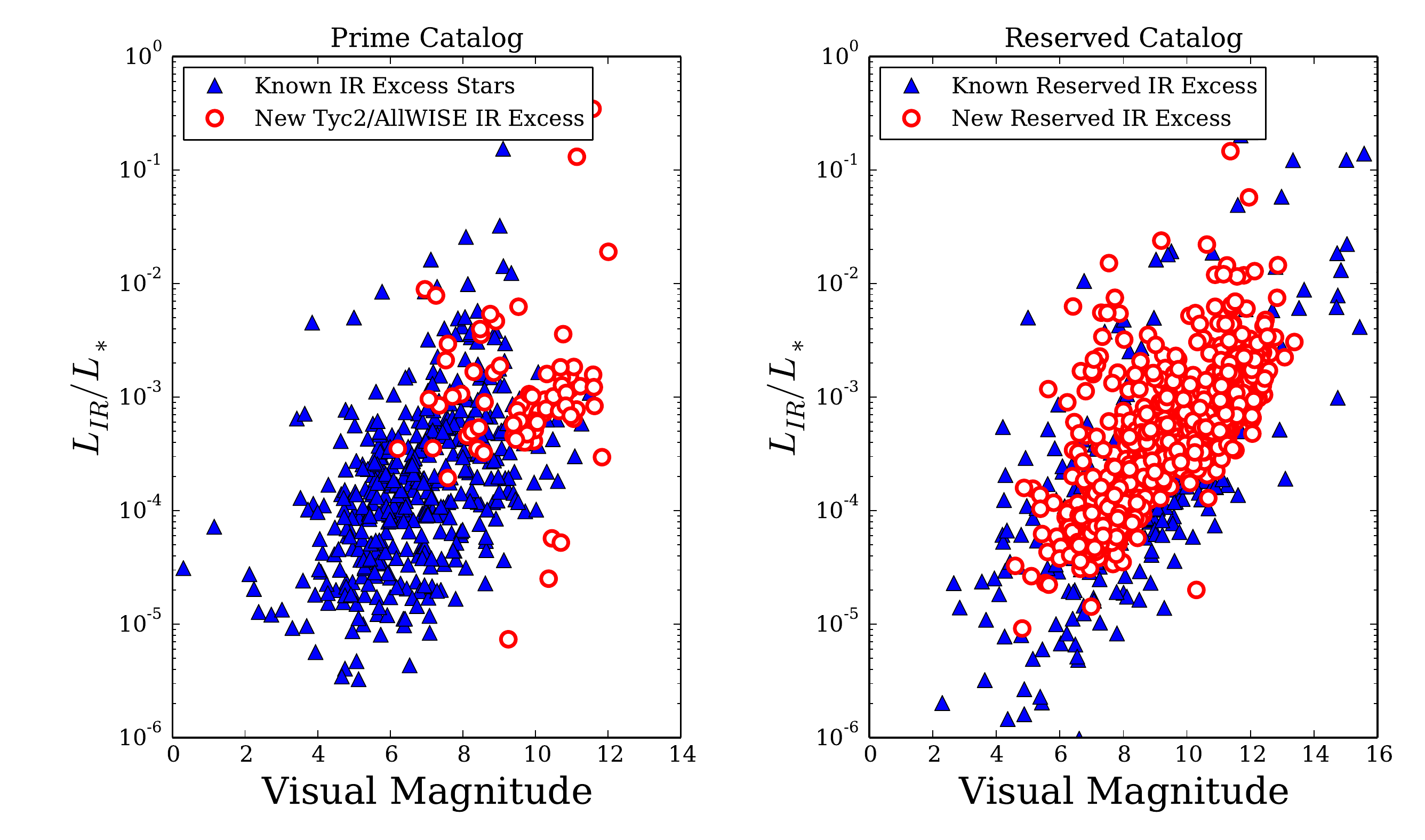}
\protect\caption{Visual magnitude versus the fractional
dust luminosity for the Prime IR excess stars (left) and the Reserved Catalog (right). 
This plot demonstrates that our new Prime IR excess stars extend to fainter magnitudes 
as the sensitivity of instruments has improved. The Reserved catalog covers the same range of magnitudes. 
There are a handful of very dusty disks ($\tau>10^{-1}$) in the Reserved catalog, however, the majority of these stars 
are likely the population of remaining giants in the sample based on an SED distance within 15 pc (see Section \ref{distcut}).}
\label{vmag_tau}
\end{centering}
\end{figure}
\end{center}

To compare our new Tycho-2/AllWISE IR excess search with the literature, 
we plot the fractional dust luminosity against the Johnson visual magnitude in Figure \ref{vmag_tau}.
Figure \ref{vmag_tau} shows our study extends the number of faint stars detected with IR excess.  
The figure also displays a trend in which these fainter magnitude stars have a higher value of $\tau$. 
This trend is indicative of later spectral types (K and M) and confirms the observational bias of our survey.
The Reserved catalog (right panel of Figure \ref{vmag_tau}) 
displays the same trend. 

Because there are a number of IR excess stars in our catalog with significantly
dusty disks ($\tau>10^{-2}$), we will take a moment just to mention these
Prime catalog targets and some of the associated parameters.

Three stars from the Prime catalog are well-known IR excess sources: 
HD 98800 (HIP 55505), BD+20 307 (HIP 8920), and HD 141569 (HIP 77542). 
These stars are within 100 pc and our SED parameters agree with literature values. 
In particular, HD 98800 was first discovered by \citet{walker1988} and later qualified to be a 
member of the TW Hydrae association \citep{zuckermansong2004b}, thereby implying the system has 
an age of roughly 8 Myr but still presents an inner region cleared of gas and dust \citep{dent2013}. 
BD+20 307 was first discovered by \citet{song2005} and then a thorough follow up of mid-IR photometry 
revealed strong silicate features \citep{weinberger2011}.  The rarity of having a significant amount 
of warm dust ($>$ 120 K) at $>$ 1 Gyr is indicative of a recent, large collisional event \citep{song2005}. 
Finally, HD 141569 is a Herbig Ae star although the nature of the IR excess is still under debate as 
to whether the disk is more debris disk-like even with the detection of gas in the disk.  Either way, the 
disk was confirmed by \citet{weinberger1999} using the \emph{Hubble Space Telescope} and so remains one 
of a few dozen stars which has been resolved through imaging the scattered light.

Besides the very well-known stars, a few other stars have been studied by \citet{chen2011} and 
\citet{olofsson2013}.  Chen et al. (2006, 2011) reported on the IR excess around HD 146897 (HIP 79977) and 
HD 129590 (HIP 72070). These solar-type stars are on the edge of our distance restriction but with values of 
fractional dust luminosity greater than $10^{-2}$ making them significantly dusty compared to the 
rest of our Prime catalog stars. In addition, HD 129590 has IRS spectroscopy 
\citep{chen2014} which confirms the amount of dust 
around this star and the use of a two component disk model fit to the dust.
HD 113766 is also best fit with two separate dust temperatures (500 K, 230 K) and \citet{olofsson2013} 
performed a full mineralogical investigation into the composition of grains in these disks.

We will address the remaining stars individually since the presentation of the IR excess 
for these sources are new detections.

\begin{center}
\begin{figure}
\begin{centering}
\begin{tabular}{cc}
\includegraphics[width=0.5\textwidth]{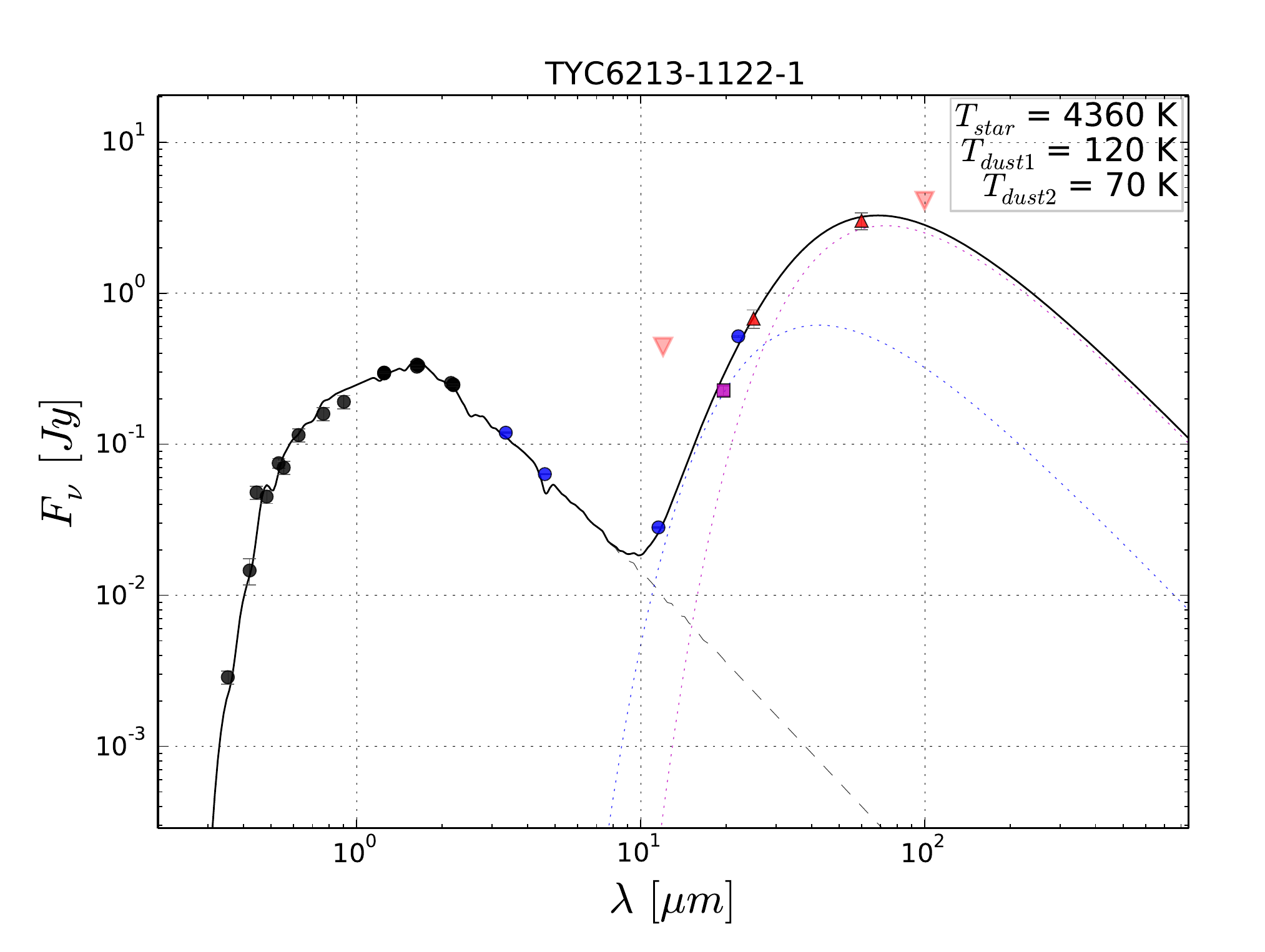} &
\includegraphics[width=0.5\textwidth]{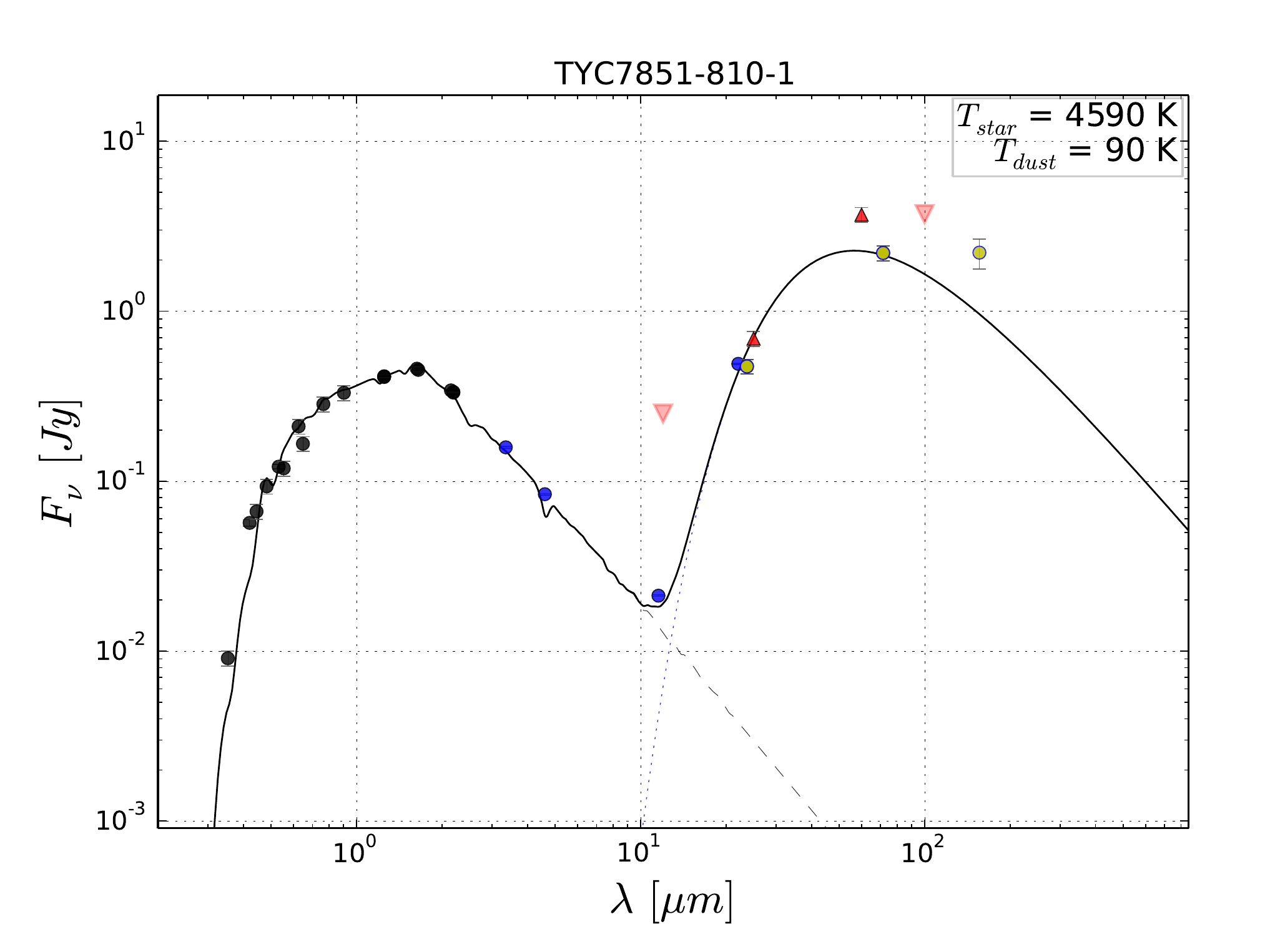} \\
\includegraphics[width=0.5\textwidth]{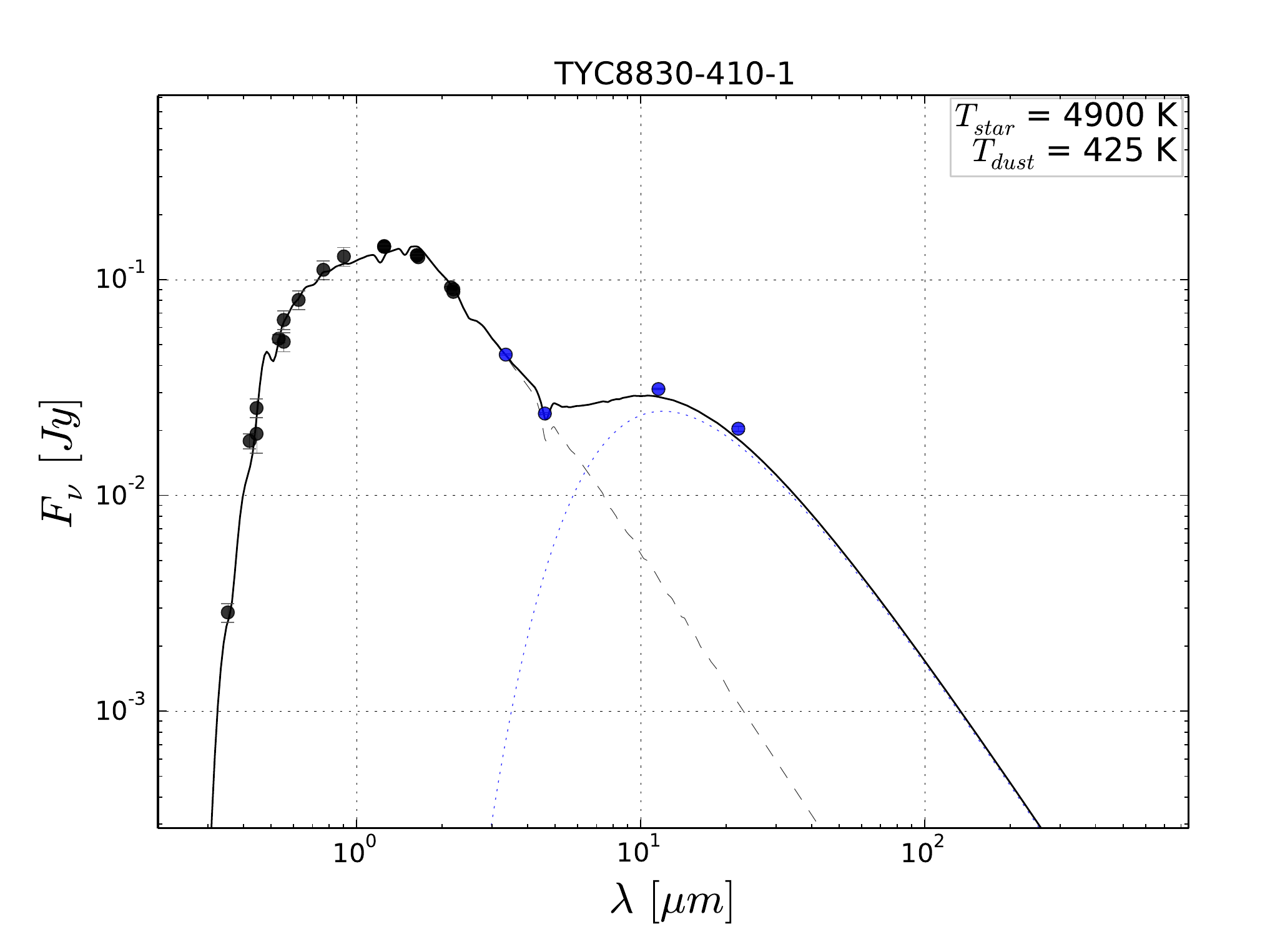} & \\
\end{tabular}
\end{centering}
\protect\caption{\label{hightau}The newly discovered, dustiest disks in the Prime IR excess catalog.
All of these targets display a fractional dust luminosity $>10^{-2}$.  
The photometry plotted in each SED includes (not necessarily in every case)
 Johnson B, V (black circles), Tycho-2 B$_{T}$, V$_{T}$ (black circles), 
\emph{Sloan} g', r', i', z' (black circles), 2MASS J, H, K (black circles), WISE W1, W2, W3, and W4 (blue circles), 
Akari 9 and 18$\mu$m (magenta squares), MIPS 24 and 70$\mu$m (yellow circles), and 
IRAS 12, 25, 60 and 100$\mu$m (red triangles) from the optical to the far-IR.  
A description of each star can be found in Section \ref{lir_lstar}.}
\end{figure}
\end{center}

\emph{TYC 6213-1122-1: }This new Tycho-2 IR excess star currently does not have an
entry in SIMBAD. It was best fit with a stellar temperature of 4370 K making it a spectral type
of $\sim$K6. The projected SED distance places this star 60.5 pc
from the Earth. In combination with its position at 16 hours right
ascension and a declination of -21 degrees, we hypothesize this star
may be a new member of the Upper Scorpius-Centaurus region. 
Figure \ref{hightau} displays the SED for TYC 6213-1122-1.
The youth of this region ($\sim$5 Myr; \citealt{preibisch1999} or $\sim$11 Myr; \citealt{pecaut2012}) 
means the large dust emission, $\tau=23.2\times10^{-2}$, may in part be attributed
to young age, however, optical spectroscopy is necessary to confirm the age of this system. 

\emph{TYC 7851-810-1: }This star is another new discovery with a similarly
dusty disk to TYC 6213-1122-1 and a similar sky
position (R.A. of 16 hours, Dec. of -38 degrees) possibly belonging to the Scorpius-Centaurus
region. A recent study by \citet{merin2008} using the \emph{Spitzer
}c2d survey which studied ``From Molecular Cores to Planet-Forming
Disks'', found this object near the ScoCen region and using the MIPS
instrument at 24 and 70$\mu$m qualified the disk to be a young, Class
II circumstellar disk. However, based on the shape of the SED,
the disk shows a large inner cleared cavity and without information regarding the gas content 
of the disk, we are led to believe this object is definitely undergoing an interesting transient phase 
regardless of its age. 
This object will be discussed further in T. Cotten et al. (2016; in prep). 
The SED provides a stellar temperature of 4590K ($\sim$K6 spectral
type), $\tau=13.1\times10^{-2}$, and a distance of 57.6 pc. 
From Figure \ref{hightau}, 
the IR excess is corroborated by photometry at 
AllWISE W3 and W4, IRAS 25, 60, and 100$\mu$m, \emph{Spitzer} MIPS 24, 70, and 160$\mu$m.

\emph{TYC8830-410-1: }This star does not have any available information or references
from SIMBAD. It is best fit using a 4900 K stellar temperature and so we assign it a spectral type of K3. 
The dust displays a luminosity of $\tau=1.90\times10^{-2}$ in the warm inner region near the star and is best fit 
using a single disk with a dust temperature of 425 K. Our SED returned a distance of 120 pc for this source. Based on 
the evidence we have for this star, the IR excess is indicative of the rare case of warm debris such as with BD+20 307. 
Based on the knowledge that silicates are expected to emit around 10$\mu$m, 
the peak of this star's IR excess at W3 (see Figure \ref{hightau})
could indicate a transient event which is generating the destruction of planetesimals, comets, and/or asteroids. 
Current ongoing observations using SOFIA (Stratospheric Observatory for Infrared Astronomy; \citealt{young2012}; 
cycle 3 \& 4 programs 03\_0099, 04\_0126, 04\_0130) should provide 
more evidence regarding the nature of the dust.

\subsection{Dust Temperatures and Disk Radii}\label{dusttemprad}

\begin{center}
\begin{figure}
\begin{centering}
\begin{tabular}{c}
\includegraphics[height=3in]{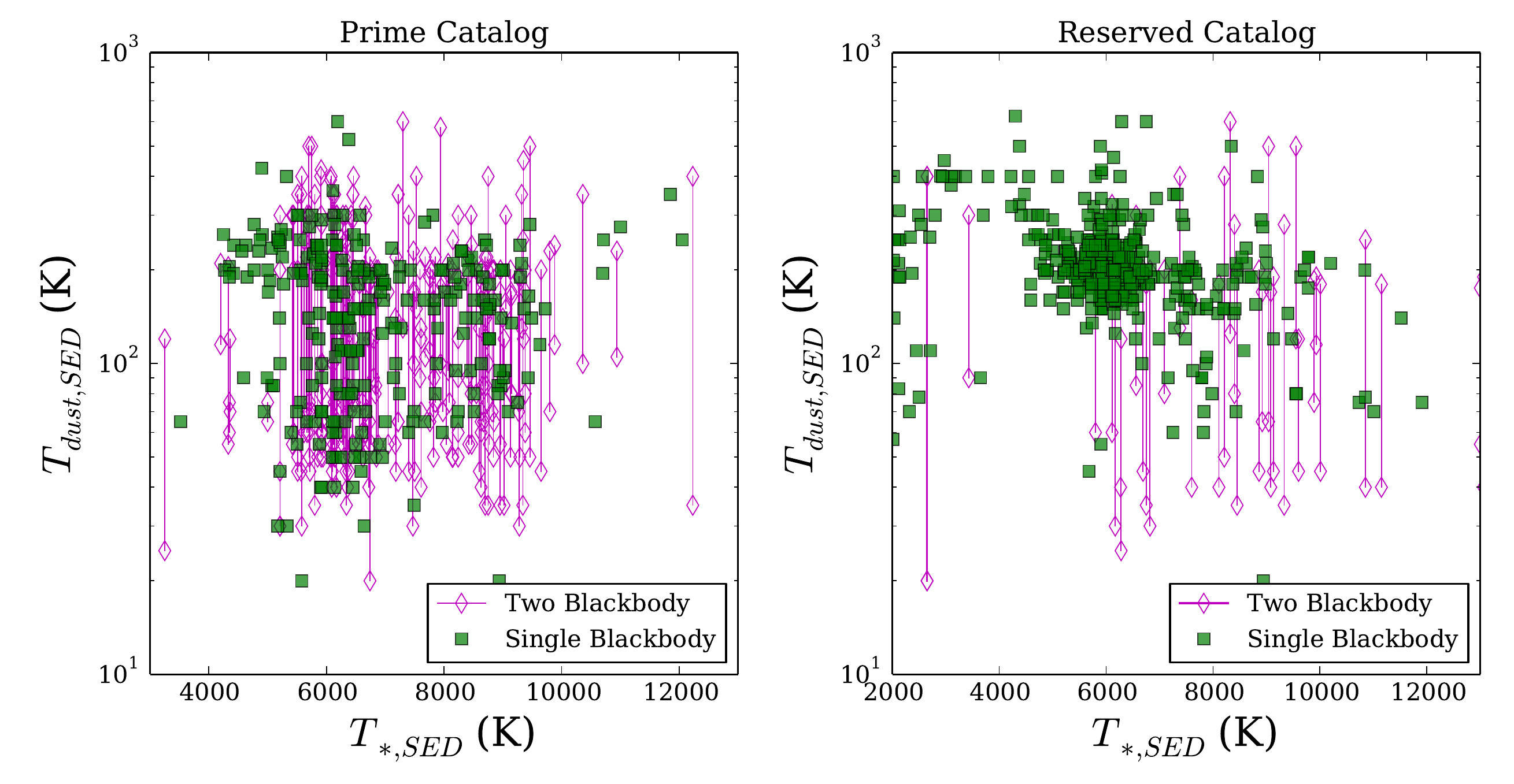} \\ 
\includegraphics[height=3in]{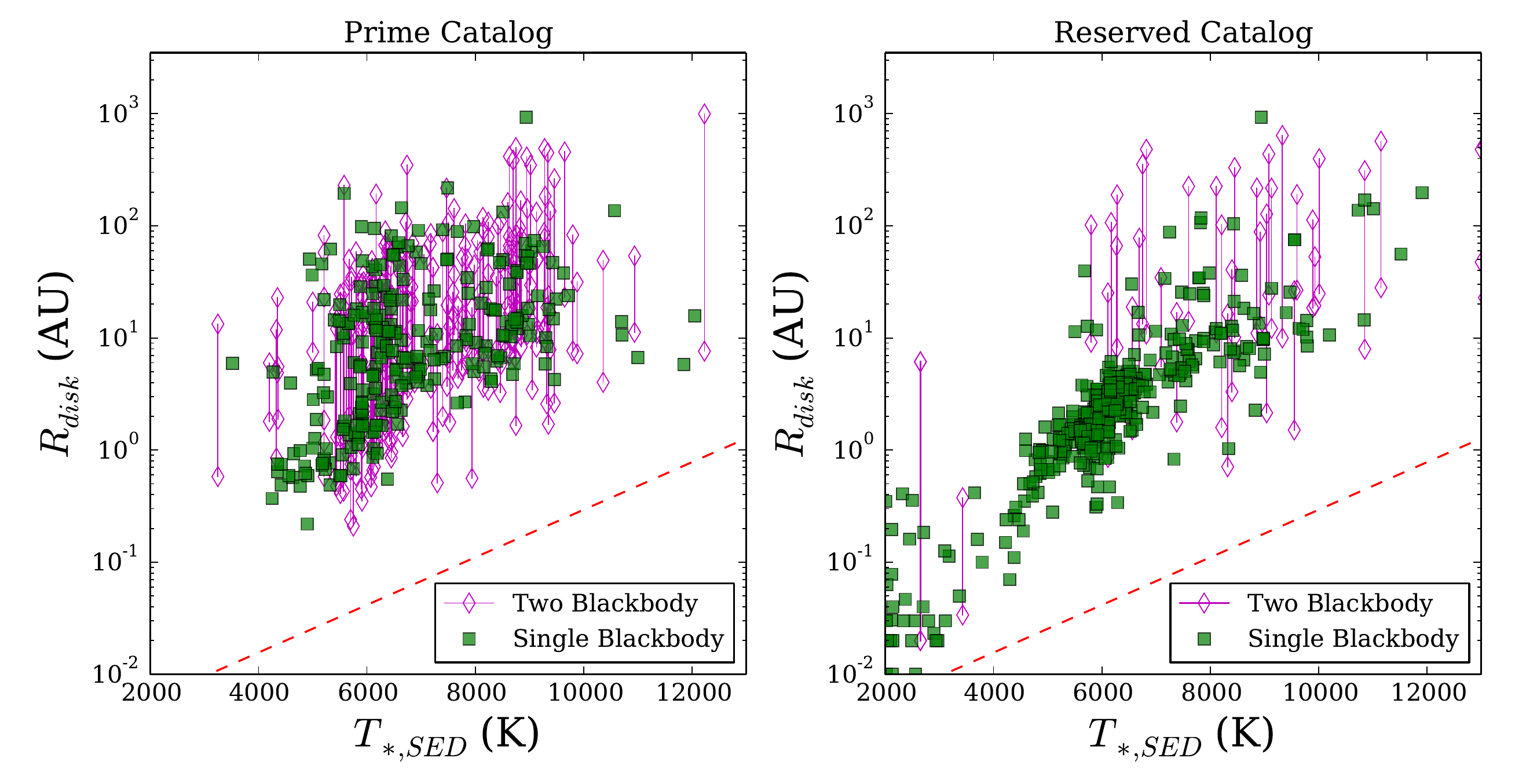} \\
\end{tabular}
\end{centering}
\protect\caption{\label{tstar_dust}Top: Dust temperature for the range 
of spectral types included in the Prime IR excess catalog (\emph{Left}; Table \ref{highfidelity}) and 
the Reserved IR excess catalog (\emph{Right}; Table \ref{reserved}).
The filled squares correspond to single blackbody dust fit. The unfilled 
diamonds are the IR excess stars that are best fit with two blackbody 
fits and the line connects the two. The sample of the Reserved catalog that is plotted 
display excess at more than one passband as without this criteria, the disk fitting procedure 
remains unconstrained.
\newline Bottom: Disk radius in
AU for the single and two blackbody fits compared the spectral type
of the star using the best fit SED temperature for the Prime catalog (\emph{Left}) 
and the Reserved catalog (\emph{Right}). The dashed red line indicates the 
dust sublimation radius for silicate grains behaving as blackbodies at a 
sublimation temperature of 1500 K (\citealt{moromartin2013}).}
\end{figure}
\end{center}

The parameters of the dust fits can be used to derive an orbital dust radius ($R_{disk}$) as in \citet{rhee2007}.  
This implies that $R_{disk} (R_{\sun}) = (\frac{R_{*,SED}}{2})(\frac{T_{*,SED}}{T_{disk}})^2$ which is 
then converted into AU.  
If the true dust grains are smaller than the model blackbody grains, they would be
located further away than blackbody grains.
Figure \ref{tstar_dust} displays the dust temperature and radius of disk. 
The symbol shapes distinguish between a single or two-component disk fit by using a connecting solid line. 
Over one third of the Prime IR excess stars have disks ($\sim37\%$) are best fit using two blackbody curves. 
The left plot displays the Prime catalog stars while the right shows the 
Reserved catalog restricted to stars that have more than one IR excess passband. 
These top plots show no recognizable trend for spectral
type and dust temperature which directly agrees with the reports mentioned in the 
review article by \citet{matthews2014}. Thus, we conclude that the luminosity and mass of the 
host star does not predict the temperature expected for the micron-sized grains creating the IR excess. 
However, the dust temperature is not the complete picture and is intimately linked to the disk radius.

Considering the results shown in the top panels of Figure \ref{tstar_dust}, 
the bottom panels confirm a trend of late-type stars having the smallest disk radii.
If all spectral types sustain disks of similar dust temperatures, then the late-type stars would need 
to have dust closer to the star to reradiate IR excess at the same temperature as early-type stars.
In addition, the estimated silicate grain sublimation range ($\sim$1500 K estimated 
by \citealt{moromartin2013}) is plotted as the red dashed line in Figure \ref{tstar_dust}. 
Because dust grains will be sublimated inside of this range, we do not see any IR excess stars with disk size 
smaller than this in our catalogs. We note, however,
that for a given stellar spectral type, the smallest disk size is at least an order of
magnitude larger than the dust sublimation limit. This implies that either grains far from
the sublimation distance are efficiently removed by other mechanisms or 
that hot disks are not identifiable observationally because of the various limits (e.g., photospheric flux
estimate, poor photometric measurement of excess emission, etc.).

\begin{center}
\begin{figure}
\begin{centering}
\includegraphics[width=5in]{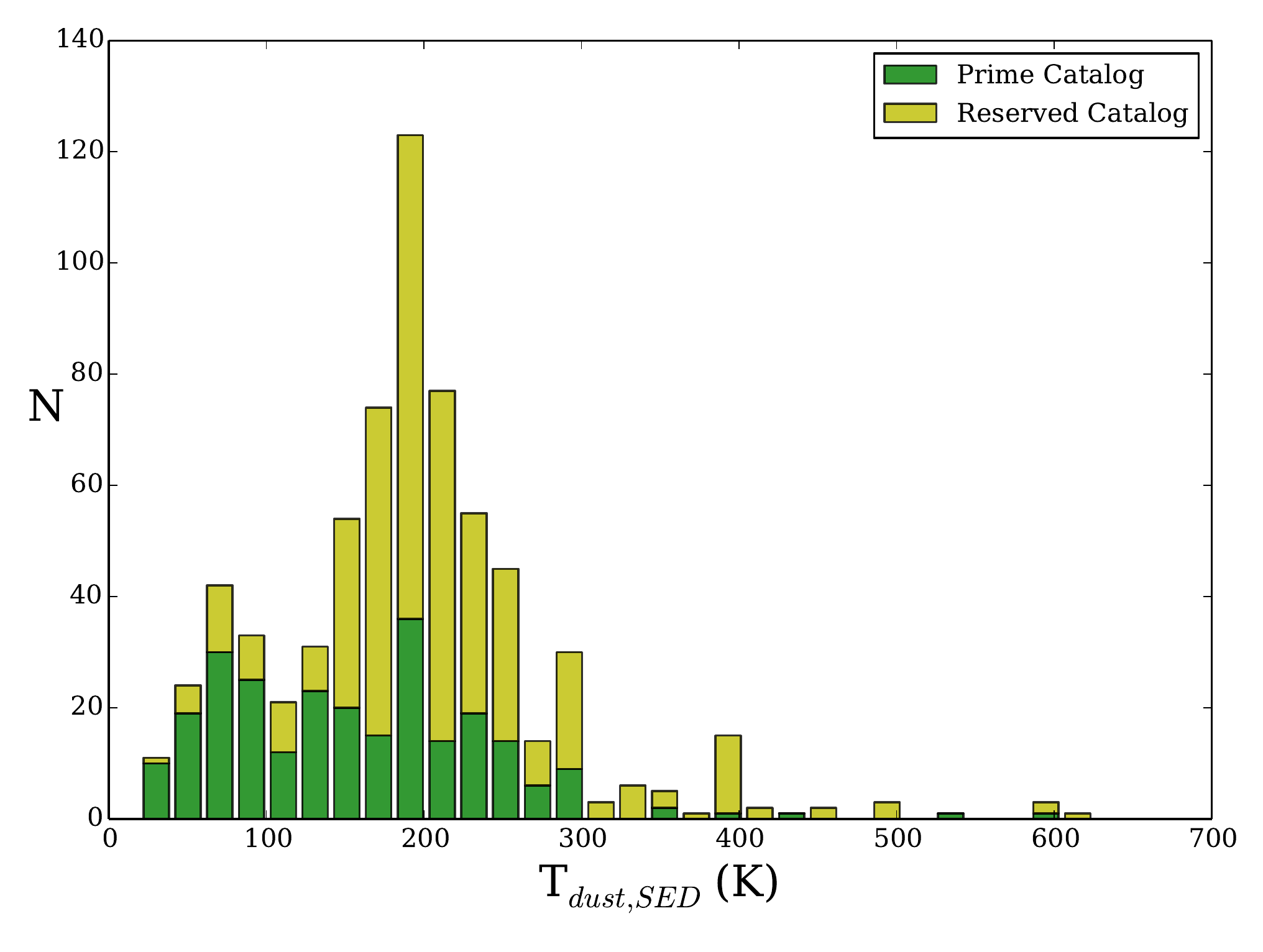}
\end{centering}
\protect\caption{\label{tdusthist}Distribution of dust temperatures for stars best fit 
using a single blackbody fit in the  
Prime and Reserved catalogs. Only stars with multiple passbands that display IR excess are included 
in this figure. }
\end{figure}
\end{center}

The dust temperature for the stars fit with a single dust temperature using multiple passbands of IR excess 
from the Prime and Reserved catalogs is shown in Figure \ref{tdusthist}. 
The distribution of dust temperatures appears to be evenly spread with a peak around 200 K. 
This peak mostly reflects the IR excess stars whose excess begins around W3 and peaks in the mid-IR.  
Since many of these disks are new discoveries using AllWISE and may not have far-IR photometry, our distribution 
may be biased in that many disks may only show the warm tail end of the dust disk. 
Separately, previous discoveries of IR excess make up the majority of the  population of disks with cooler dust temperatures 
around 80 K. This is due to the fact that the known IR excess stars were mainly discovered 
at far-IR wavelengths where the excess begins at or after W4. 

\begin{center}
\begin{figure}
\begin{centering}
\includegraphics[width=5in]{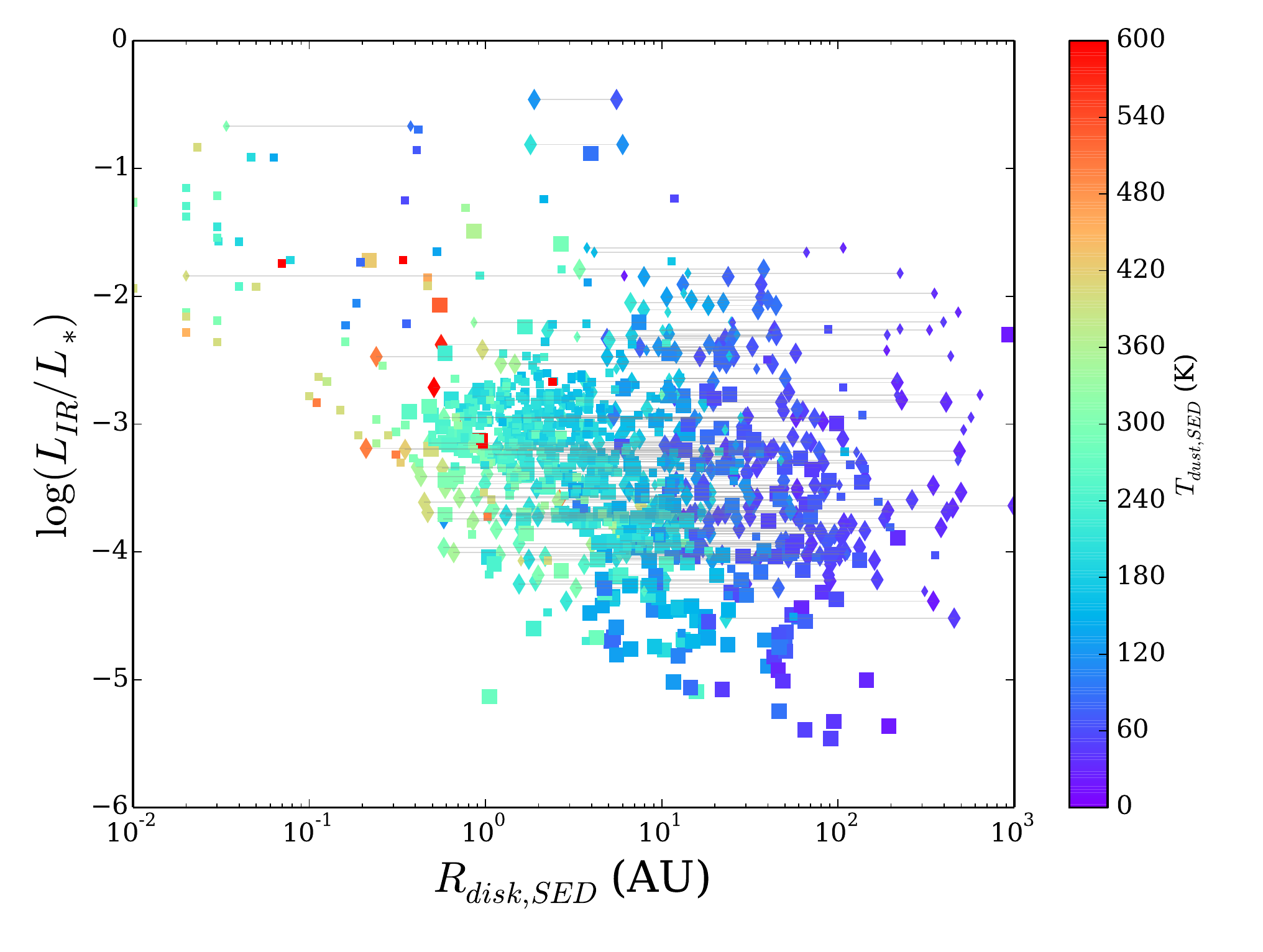}
\protect\caption{Disk radius in AU versus the fractional dust
luminosity ($\tau$) for the Prime (large symbols) and Reserved (small symbols) 
IR excess catalogs. The squares correspond to single blackbody dust fit. The diamonds are
the stars that are best fit with two blackbody fits and
the line connects the two. The color corresponds to the temperature
of the dust.}
\label{tau_dust}
\end{centering}
\end{figure}
\end{center}

Figure \ref{tau_dust} compares the fractional dust luminosity to the dust radius.  
The temperatures of the single blackbody fits and the two components of the dust are plotted
as the color in the figure. One may expect that an IR excess star
with multiple disk components would exhibit a higher fractional dust luminosity, but Figure
\ref{tau_dust} does not display such a trend. 
Additionally, this figure confirms that the warm disk components are closer to the star, while the 
cooler components are further away. 
This plot does display an apparent void of low fractional dust luminosity and small disk radius. 
As referenced earlier, this deficit of stars is likely an observational bias due to the difficulty in assessing IR 
excess from the photosphere at wavelengths shorter than 10$\mu$m which may produce a collection of marginal, warm IR 
excess stars. 

\begin{center}
\begin{figure}
\begin{centering}
\includegraphics[width=5in]{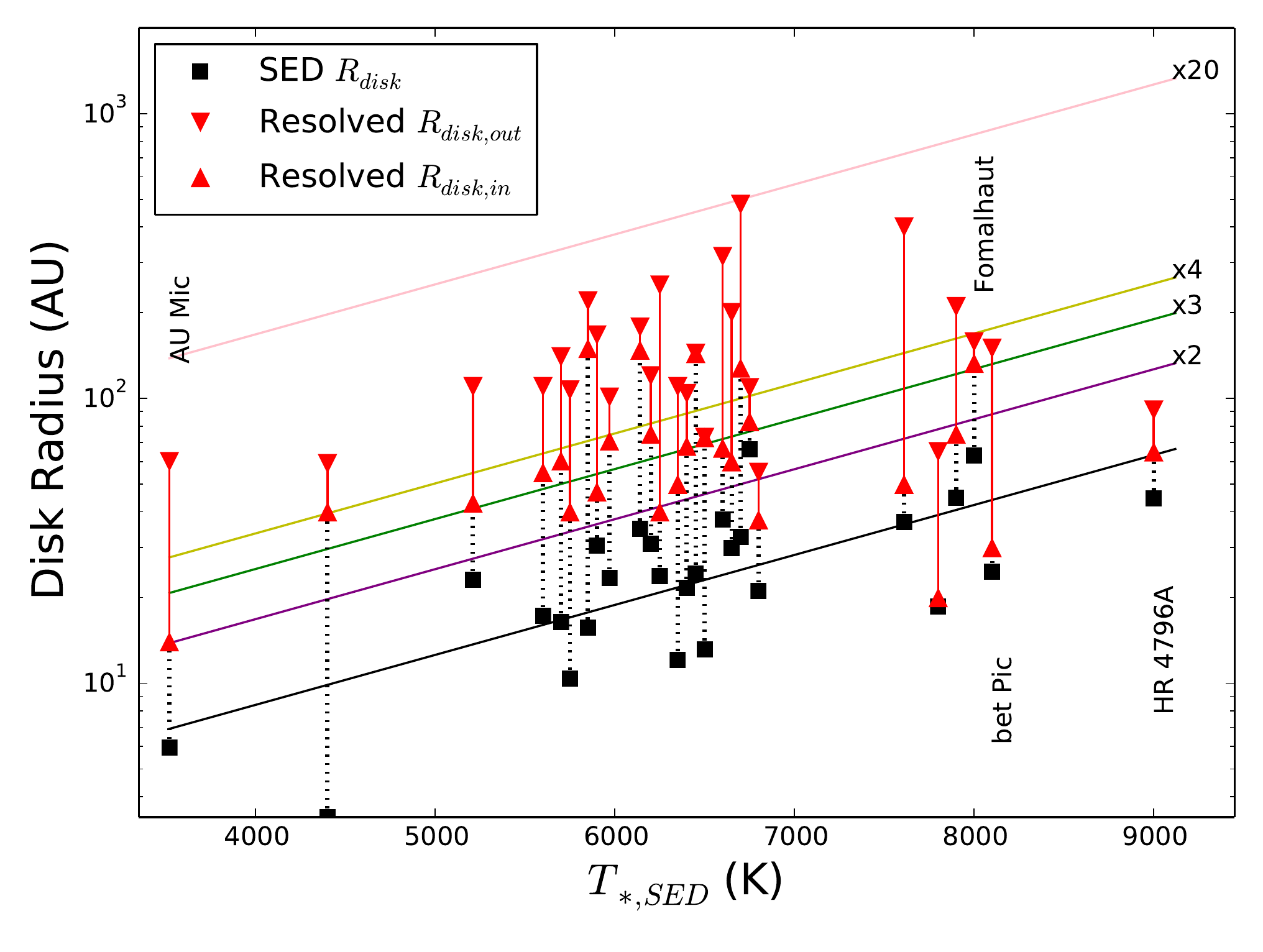}
\protect\caption{Comparison of the disk radius predicted by the SED blackbody model fit 
and the radius resolved through scattered light. The black squares are the SED disk radius in AU and the 
red triangles connected with the solid line are the extended dust range from inner to outer radius in AU. 
The names of some of the most well-known debris disks are shown. The unnamed stars are flagged in Tables 
\ref{highfidelity} and \ref{reserved}. The solid black line 
indicates the fit to the SED disk radius (squares) and the consecutive lines indicate the amount of increase 
to that fit.  The purple, green, yellow, and pink line shows twice, three times, four times, 
and 20 times the original fit, respectively.}
\label{resolved}
\end{centering}
\end{figure}
\end{center}

Studies of thermally resolved disks by \citet{booth2013} and \citet{rodriguez2012} 
found that the predicted disk radius from the SED blackbody fit was 1-5 times smaller 
then the size of the disk resolved in thermal imaging. 
They conclude from their results that the dust grains studied using different techniques 
must be of different sizes and compositions and in particular, thermal imaging better traces the larger grains.
Scattered light imaging probes a small grain size (sub-micron) and yet this ratio has been used to compare this method to 
the SED disk radius.
In Figure \ref{resolved}, we display the predicted disk radius from our blackbody SED fit 
to the disk radius for stars that have been resolved through scattered light. 
There are over two dozen of these stars in our catalogs and they have been identified in the tables. 
A few of the most well-known debris disk stars have been labeled in the figure including AU Mic, Fomalhaut, and beta Pictoris. 
Figure \ref{resolved} shows that we can make a prediction regarding the relationship between different dust detection methods. 
The SED disk radii (shown as black squares) displays the trend shown by the best-fit solid black line and 
the additional lines shown in this figure are multiplicative factors of this line. 
Since many of the disks are extended, we include the inner and 
outer components of the continuous disk rather than an average or peak location of the dust. This plot 
demonstrates the resolved inner disk boundary ranges from 1 to 4 times the blackbody disk radius 
in agreement with the results from \citet{rodriguez2012} and \citet{booth2013}. 
Further, we extend this relationship to the outer edge of the dust disks and estimate 
the outer edge of the circumstellar material can extend to nearly 20 times the blackbody dust radius.  

Biases between the disks resolved through scattered light and SED blackbody modeling complicate 
the comparison made in Figure \ref{resolved}. For many disks imaged in scattered light, the inner portion of the disk suffers 
from self-subtraction due to the large stellar flux and so an inner rim is difficult to measure. On the far side, 
the dust grains tend to fade out of view of the central star's flux at large distances and so an outer rim is 
difficult to determine as well. Further, the inclination angle of the disk presents difficulties in determining an 
inner and outer edge, especially if the disk is edge-on like AU Mic. In these cases, the inner rim of the dust is 
approximated by the closest the scattered light technique can probe to the central star. The comparison shown in 
Figure \ref{resolved} is still useful as it can be extrapolated to the entire Prime catalog and used as a 
method to identify targets for future observations and improve predictions regarding the expected dust location.

\section{Conclusion}\label{conclude}

We have conducted an extensive search and collection of IR excess
stars from two sources: a new Tycho-2 cross-correlation with the AllWISE
catalog and a literature search for previously claimed IR excess 
stars. We recognize a need for a saturation correction in AllWISE W2 and develop 
a simple polynomial correction. 
Using a SED fitting algorithm, we determine the amount of excess above the photosphere comparing
the measured and the photospheric fluxes. We select new IR excess candidates
by requiring an excess of $>3\sigma$ or $>5\sigma$ in either W3 or W4 
and infrared bright in either W3 or W4 ($>$10 mJy). 
Further, extensive analysis of the AllWISE images removed a large number of
false-positives. We then implement a series of criteria involving
brightness, number of passbands showing excess, and distance in order
to ensure a sample of the highest fidelity IR excess stars displaying an SED indicative 
of a post-protoplanetary disk.
For stars lacking an accurate \emph{Hipparcos
}parallax, we perform an estimate of distance using the best fit
SED temperature and extrapolating a typical stellar radius from a composite isochrone. 
This is similar to a photometric distance, but with
improved accuracy using all the available photometric data from the
optical to infrared wavelengths. Finally, the inclusion of catalogs
such as \emph{Akari}, IRAS, AllWISE, \emph{Spitzer }MIPS, IRAS, and \emph{Herschel
} PACS and SPIRE improves the reliability of the Prime catalog, especially
when the measurements corroborate and reduce the degeneracy
of the SED fitting algorithm.

Specifically focusing on our Prime sample of $\sim$500 nearby ($<120$ pc) IR excess 
stars, we have compiled the largest, most reliable IR excess 
star catalog to date. In addition, the new Tycho-2/AllWISE search
for IR excess increased the known, reliable IR excess sample by $\sim$20\%. 
Moreover, a few IR excess stars appear to be new, extremely dusty systems
requiring follow-up observations to better understand the evolution
of the dust and/or the transient nature of the dust.
The literature IR excess sample represents mainly 
the cold, less dusty disks, while the newly discovered Tycho-2/AllWISE IR excess stars 
are typically warmer and more dusty. 
Considering the Reserved sample, this study more than doubled the number of known IR excess stars 
and these stars are maintained for more sensitive future surveys.

We offer a discussion of the relationship between the dust parameters and stellar parameters
obtained during the SED fitting procedure and a portrayal of
many two disk systems which span the entire spectral type range. 
Our findings affirm that a two component dust disk does not suggest any particular
stellar or dust temperature, but also that the activity which generates the dust around these stars 
can be assumed to be analogous to either the Asteroid or the Kuiper belt and operates regardless of the luminosity 
or mass of the host star. 
Future work should strive for a complete catalog of sub millimeter data which has been shown 
to be more suggestive of the true amount of dust in these systems and thus, provide a better suggestion
of the mass of dust in each system. In addition, we investigate the relationship between the disk radius 
assumed using a blackbody disk model and the disk radius resolved using scattered light. Since scattered light 
reveals the actual location of dust in the disk, the SED disk radius can be used to indicate that 
the true inner disk radius is roughly four times larger and the outer disk radius is twenty times larger than predicted 
by the SED.

\begin{acknowledgments}
We thank the referee for their very helpful comments and suggestions that have improved this manuscript. 
Support for this work was provided by NASA (NNN12AA01C, NAS2-97001) through an award issued by JPL/Caltech and USRA as well as 
partially supported by a grant issued by SOFIA and USRA to UGA.
The authors would also like to acknowledge J. Lee, A. Schneider, and L. Sgro for their helpful suggestions.
This study has made extensive use of the NASA/ IPAC Infrared Science Archive, 
data products from the Wide-field Infrared Survey Explorer, 
the \emph{Herschel} Science archive, the \emph{Spitzer} Heritage Archive, SIMBAD, and VizieR (operated at CDS).
\end{acknowledgments}

\input{bib_cottensong2016arXiv}

\clearpage

\onecolumn

\pagebreak
\clearpage

\begin{center}
\begin{deluxetable}{|c|c|c|c|c|}
\tabletypesize{\scriptsize}
\tablewidth{0pc}
\tablecaption{Completed Infrared Space Mission Instruments.\label{irscopes}}
\tablehead{
\colhead{Instrument} & \colhead{Filter Wavelength}  & \colhead{Ang Res} & \colhead{Yr of Launch} & \colhead{Comments}\\
\colhead{} & \colhead{($\mu m$)} & \colhead{(arcsec)} & \colhead{} & \colhead{} \\
}
\startdata
IRAS & 12, 25, 60, 100 & \multicolumn{1}{c|}{30 - 120} & 1983 & all-sky, 96\%, 250,000 sources\\
ISO & 2.5 - 240 & 1.5 - 90 & 1995 & 30,000 pointed obs.\\
\emph{Spitzer }MIPS & 24, 70, 160 & 6 - 40 & 2003 & $\sim$42 million pointed obs.\\
Akari & 9, 18 & \textasciitilde{}60 & 2006 & all-sky, 870,000 sources\\
\emph{Herschel }PACS & 70, 100, 160 & 5 - 13 & 2008 & $\sim$40,000 pointed obs.\\
\emph{Herschel }SPIRE & 250, 350, 500 & 18 - 36 & 2008 & $\sim$40,000 pointed obs.\\
WISE & 3.5, 4.6, 12, 22 & 6.1 - 12 & 2009 & all-sky, $>$500 million sources\\
AllWISE & 3.5, 4.6, 12, 22 & 6.1 - 12 & 2009 & all-sky, $>$740 million sources\\
\enddata
\end{deluxetable}
\end{center}

\begin{center}
\begin{deluxetable}{|cc|cc|cc|}
\tabletypesize{\scriptsize}
\tablewidth{0pt}
\tablecaption{Summary of References used in literature search for previously claimed IR excess stars.\label{refs}}
\tablehead{
\colhead{Author} & \colhead{Year} & 
\colhead{Author} & \colhead{Year} &
\colhead{Author} & \colhead{Year} \\
}
\startdata
 Aumann et al.            &1984&  
 Sadakane \& Nishida              &1986&  
 Jura                        &1991\\  
 Patten \& Willson                &1991&  
 Oudmaijer et al.         &1992&  
 Mannings \& Barlow               &1998\\  
 Decin et al.                &2000&  
 Song et al.                 &2001&  
 Spangler et al.             &2001\\  
 Sylvester et al.        &2001&  
 Laureijs et al.         &2002&  
 Song et al.                 &2002\\  
 Weinberger et al.       &2003& 
 Wyatt et al.            &2003& 
 Gorlova et al.             &2004\\  
 Greaves et al.          &2004&  
 Jura et al.                &2004&  
 Liu et al.              &2004\\  
 Metchev et al.           &2004&  
 Zuckerman \& Song        &2004&  
 Zuckerman et al.           &2004\\  
 Beichman et al.         &2005&  
 Chen et al.              &2005&  
 Chen et al.             &2005\\  
 Greaves et al.           &2005&  
 Kim et al.               &2005&  
 Low et al.           &2005\\  
 Najita et al.               &2005&  
 Rieke et al.            &2005&  
 Song et al.                &2005\\  
 Su et al.            &2005&  
 Beichman et al.         &2006& 
 Beichman et al.         &2006\\  
 Bryden et al.              &2006&  
 Bryden et al.              &2006&  
 Carpenter et al.        &2006\\  
 Chen et al.              &2006&  
 Gorlova et al.             &2006&  
 Goto et al.                &2006\\  
 Hernandez et al.           &2006&  
 Hines et al.             &2006&  
 Kalas et al.               &2006\\  
 Lestrade et al.         &2006&  
 Moor et al.                &2006&  
 Riaz et al.                &2006\\  
 Silverstone et al.      &2006&  
 Smith et al.            &2006&  
 Su et al.            &2006\\  
 Williams \& Andrews     &2006&  
 Cieza et al.            &2007&  
 Gautier et al.      &2007\\  
 Gorlova et al.              &2007&  
 Guieu et al.               &2007&  
 Hernandez et al.           &2007\\  
 Kalas et al.               &2007&  
 Kalas et al.               &2007&  
 Lafreniere et al.          &2007\\ 
 Lisse et al.            &2007&  
 Luhman et al.           &2007&  
 Matthews et al.         &2007\\  
 Matthews et al.         &2007&
 Moerchen et al.         &2007&  
 Moerchen et al.          &2007\\  
 Moro-Martin et al.         &2007&  
 Rhee et al.                     &2007& 
 Rhee et al.             &2007\\  
 Siegler et al.             &2007&  
 Trilling et al.          &2007&  
 Wyatt et al.            &2007\\  
 Wyatt et al.            &2007&  
 Absil et al.               &2008&  
 Brown et al.            &2008\\  
 Chen et al.             &2008&  
 Cieza et al.            &2008&  
 Gautier et al.      &2008\\ 
 Hernandez et al.           &2008&  
 Hillenbrand et al.       &2008&  
 Kastner et al.          &2008\\  
 Marois et al.              &2008&  
 Merin et al.               &2008&  
 Meyer et al.            &2008\\  
 Rebull et al.           &2008&  
 Rhee et al.             &2008&  
 Roberge \& Weinberger      &2008\\ 
 Smith et al.                &2008&  
 Su et al.            &2008&  
 Trilling et al.         &2008\\ 
 Weinberger               &2008&  
 Wyatt                    &2008&  
 Akeson et al.           &2009\\  
 Balog et al.               &2009&  
 Booth et al.               &2009&  
 Brown et al.            &2009\\  
 Bryden et al.              &2009&  
 Carpenter et al.        &2009&  
 Fujiwara et al.            &2009\\  
 Fujiwara et al.            &2009&  
 Gaspar et al.              &2009&  
 Greaves et al.          &2009\\  
 Gutermuth et al.        &2009&  
 Hernandez et al.           &2009&  
 Kospal et al.              &2009\\  
 Lawler et al.           &2009&  
 Lestrade et al.         &2009&  
 Maness et al.           &2009\\  
 Melis et al.               &2009&  
 Metchev et al.             &2009&  
 Moor et al.                &2009\\  
 Morales et al.      &2009&  
 Nilsson et al.             &2009&  
 Plavchan et al.         &2009\\  
 Roccatagliata et al.       &2009&  
 Schneider et al.           &2009&  
 Sicilia-Aguilar et al.      &2009\\  
 Smith et al.               &2009&  
 Smith et al.               &2009&  
 Su et al.            &2009\\ 
 Tanner et al.              &2009&  
 Ardila et al.           &2010&  
 Bonsor et al.              &2010\\  
 Buenzli et al.             &2010&  
 Cieza et al.            &2010&  
 Duchene                     &2010\\  
 Eiroa et al.               &2010&  
 Gizis                    &2010&  
 Greaves et al.           &2010\\  
 Grinin et al.           &2010&  
 Kastner et al.          &2010&  
 Kennedy et al.            &2010\\  
 Koerner et al.          &2010& 
 Krivov et al.           &2010&  
 Lagrange et al.        &2010\\  
 Liseau et al.                &2010&  
 Matthews et al.         &2010&  
 Melis et al.               &2010\\  
 Moerchen et al.          &2010&  
 Monin et al.            &2010&  
 Moro-Martin et al.         &2010\\  
 Nilsson et al.             &2010&  
 Rebull et al.           &2010&  
 Sierchio et al.         &2010\\  
 Smith \& Wyatt          &2010&  
 Stock et al.            &2010&  
 Thompson et al.         &2010\\  
 Vandenbussche et al.         &2010& 
 Wahhaj et al.              &2010&  
 Chen et al.             &2011\\  
 Churcher et al.           &2011&  
 Churcher et al.           &2011&  
 Currie et al.              &2011\\  
 Desidera et al.            &2011&  
 Eiroa et al.                &2011&  
 Golimowski et al.        &2011\\  
 Heng                        &2011&  
 Kennedy et al.            &2011&  
 Marshall et al.          &2011\\  
 Millan-Gabet et al.        &2011&  
 Moor et al.                &2011&  
 Morales et al.          &2011\\  
 Patience et al.            &2011&  
 Peterson et al.         &2011&  
 Smith et al.               &2011\\  
 Williams \& Cieza        &2011&  
 Wilner et al.        &2011&  
 Zuckerman et al.             &2011\\  
 Acke et al.                 &2012&  
 Avenhaus et al.            &2012&  
 Chen et al.             &2012\\  
 Cieza et al.             &2012&  
 Dahm et al.             &2012&  
 Donaldson et al.           &2012\\  
 Ertel et al.                 &2012&  
 Kennedy et al.            &2012&  
 Lawler \& Gladman           &2012\\  
 Lebreton et al.              &2012&  
 Lestrade et al.         &2012&  
 Luhman \& Mamajek  &2012\\  
 Maldonado et al.           &2012&  
 Melis et al.              &2012&  
 Meng et al.          &2012\\  
 Mizusawa et al.           &2012&  
 Morales et al.          &2012&  
 Riaz \& Gizis                &2012\\  
 Riviere-Marichalar et al.  &2012&  
 Riviere-Marichalar et al.    &2012&  
 Rodigas et al.          &2012\\  
 Rodriguez \& Zuckerman        &2012&  
 Schneider et al.           &2012&  
 Smith \& Jeffries           &2012\\  
 Urban et al.            &2012&  
 Wyatt et al.             &2012&  
 Zuckerman \& Song        &2012\\  
 Zuckerman et al.           &2012&  
 Ballering et al.            &2013&  
 Bonsor et al.               &2013\\  
 Booth et al.                 &2013&  
 Broekhoven-Fiene et al.     &2013&  
 Bulger et al.              &2013\\  
 Eiroa et al.                 &2013& 
 Fujiwara et al.            &2013&  
 Gaspar et al.           &2013\\  
 Janson et al.               &2013&  
 Mathews et al.          &2013&  
 Melis et al.              &2013\\  
 Moor et al.                &2013&  
 Morales et al.          &2013&  
 Olofsson et al.              &2013\\  
 Panic et al.               &2013&  
 Riviere-Marichalar et al    &2013&  
 Schneider et al.            &2013\\  
 Thalmann et al.            &2013&  
 Wahhaj et al.               &2013&  
 Bailey et al.              &2014\\  
 Ballering et al.            &2014&  
 Bonsor et al.              &2014&  
 Carpenter et al.        &2014\\  
 Esplin et al.           &2014&  
 Greaves et al.          &2014&  
 Greaves et al.          &2014\\  
 Kennedy et al.            &2014&  
 Kennedy et al.            &2014&
 Liu et al.               &2014\\  
 Marshall et al.         &2014&  
 Panic et al.               &2014&  
 Pawellek et al.             &2014\\  
 Ricci et al.               &2014&  
 Riviere-Marichalar et al.   &2014&  
 Rodigas et al.          &2014\\  
 Soummer et al.             &2014&  
 Thureau et al.           &2014&  
 Wittenmyer et al.       &2014\\  
 Hung et al.              &2015&  
 Jang-Condell et al.         &2015&  
 Maldonado et al.            &2015\\ 
 Moor et al.                &2015&  
 Rodigas et al.          &2015& & \\  
\enddata
\end{deluxetable}
\end{center}
\clearpage

\input{newhf}
\pagebreak

\input{newreserved}
\pagebreak

\begin{center}
\begin{deluxetable}{ll}
\tablewidth{0pt}
\tablecaption{Photometry Information for Prime and Reserved Catalog Stars.\label{phottable}}
\tablehead{
Column & Explanation \\
}
\startdata
1 &  Source identifier \\
2-8 & RA/DE (J2000) \\
9 & Catalog Membership \\
10 & SED Effective Temperature \\
11 & Stellar Radius (R$_{\odot}$) \\
12-14 & W1 measurement (Jy), Uncertainty, Predicted photospheric measurement \\
15-17 & W2 measurement (Jy), Uncertainty, Predicted photospheric measurement \\
18-20 & W3 measurement (Jy), Uncertainty, Predicted photospheric measurement \\
21-23 & W4 measurement (Jy), Uncertainty, Predicted photospheric measurement \\
24-26 & IRAS 12$\mu$m measurement (Jy), Uncertainty, Predicted photospheric measurement  \\
27-29 & IRAS 25$\mu$m measurement (Jy), Uncertainty, Predicted photospheric measurement  \\
30-32 & IRAS 60$\mu$m measurement (Jy), Uncertainty, Predicted photospheric measurement  \\
33-35 & IRAS 100$\mu$m measurement (Jy), Uncertainty, Predicted photospheric measurement  \\
36-39 & MIPS 24$\mu$m measurement (Jy), Uncertainty, Reference, Predicted photospheric measurement \\
40-43 & MIPS 70$\mu$m measurement (Jy), Uncertainty, Reference, Predicted photospheric measurement \\
44-47 & MIPS 160$\mu$m measurement (Jy), Uncertainty, Reference, Predicted photospheric measurement \\
48-51 & PACS 70$\mu$m measurement (Jy), Uncertainty, Reference, Predicted photospheric measurement \\
52-55 & PACS 100$\mu$m measurement (Jy), Uncertainty, Reference, Predicted photospheric measurement \\
56-59 & PACS 160$\mu$m measurement (Jy), Uncertainty, Reference, Predicted photospheric measurement \\
60-63 & SPIRE 250$\mu$m measurement (Jy), Uncertainty, Reference, Predicted photospheric measurement \\
64-67 & SPIRE 350$\mu$m measurement (Jy), Uncertainty, Reference, Predicted photospheric measurement \\
68-71 & SPIRE 500$\mu$m measurement (Jy), Uncertainty, Reference, Predicted photospheric measurement \\
\enddata
\newline
Table 5 is published in its entirety in the electronic edition \\
of the {\it Astrophysical Journal}.  A portion is shown here \\
for guidance regarding its form and content.
\end{deluxetable}
\end{center}

\end{document}

%% file: bib_cottensong2016arXiv.tex
\begin{footnotesize}

\end{footnotesize}

%% file: newhf.tex
{\footnotesize
\begin{landscape}
\setlength\LTleft{0pt}
\setlength\LTright{4pt}
\setlength\tabcolsep{2pt}
\begin{longtable}{ccccccccccccccc}
\caption{Prime IR Excess Stars from the Literature and Tycho-2 cross-correlation with AllWISE\label{highfidelity}}\\

\hline\noalign{\vskip 2mm}
\multicolumn{1}{c}{Name} & \multicolumn{1}{c}{R.A. \& Dec} & \multicolumn{1}{c}{\shortstack{Spectral\\Type}} & \multicolumn{1}{c}{\shortstack{$T_{*,SED}$\\ (K)}}  & \multicolumn{1}{c}{\shortstack{$R_{*,SED}$\\ ($R_{\sun}$)}} & \multicolumn{1}{c}{\shortstack{$T_{dust}$\\ (K)}} & \multicolumn{1}{c}{\shortstack{$R_{disk}$\\ (AU)}} & \multicolumn{1}{c}{\shortstack{$T_{dust2}$\\ (K)}} & \multicolumn{1}{c}{\shortstack{$R_{disk2}$\\ (AU)}} & \multicolumn{1}{c}{\shortstack{$\frac{L_{IR}}{L_{star}}$\\ ($\times10^{-4}$)}} & \multicolumn{1}{c}{\shortstack{Dist.\\ (pc)}} & \multicolumn{1}{c}{\shortstack{Num\\ Excess}} & \multicolumn{1}{c}{\shortstack{$\lambda_{start}$\\($\mu$m)}} & \multicolumn{1}{c}{\shortstack{Has\\ IRS?}} &\multicolumn{1}{c}{\shortstack{Known \\ Reference}} \\
\hline
\endfirsthead

\multicolumn{15}{c}{{\tablename} \thetable{} -- Continued} \\[0.5ex]
\hline\noalign{\vskip 2mm}
\multicolumn{1}{c}{Name} & \multicolumn{1}{c}{R.A. \& Dec} & \multicolumn{1}{c}{Sp. Type} & \multicolumn{1}{c}{$T_{*,SED}$}  & \multicolumn{1}{c}{$R_{*,SED}$} & \multicolumn{1}{c}{$T_{dust}$} & \multicolumn{1}{c}{$R_{disk}$} & \multicolumn{1}{c}{$T_{dust2}$} & \multicolumn{1}{c}{$R_{disk2}$} & \multicolumn{1}{c}{$\frac{L_{IR}}{L_{star}}$} & \multicolumn{1}{c}{Dist.} & \multicolumn{1}{c}{Num} & \multicolumn{1}{c}{$\lambda_{start}$} & \multicolumn{1}{c}{IRS?} & \multicolumn{1}{c}{Known?} \\[0.5ex] \hline
\endhead

\hline
\multicolumn{15}{l}{{Continued on Next Page\ldots}} \\
\endfoot

\\[-1.8ex]
\hline \hline
\endlastfoot
HR 9102 & 00:04:20.3 -29:16:07&A0V	& 8820	& 2.58	& 200	& 11.6	& 70	& 95.5	& 1.37	& 124.84& 2	& 22	& N 	& \citealt{su2006}\\
HD 105 & 00:05:52.6 -41:45:11&G0V	& 6070	& 1.01	& 390	& 0.5	& 50	& 34.6	& 4.53	& 39.38	& 3	& 22	& Y	& \citealt{zuckermansong2004a}\\
HD 166 & 00:06:37.1 +29:01:15&K0V	& 5700	& 0.78	& 300	& 0.6	& 70	& 12.1	& 3.56	& 13.67	& 4	& 22	& Y	& \citealt{bryden2006b}\\
HD 203 & 00:06:50.1 -23:06:27&F3V	& 6710	& 1.48	& 160	& 6.0	& -- 	& -- 	& 1.69	& 39.38	& 3	& 22	& Y	& \citealt{rebull2008}\\
HD 377$^{b}$ & 00:08:25.8 +06:37:00&G2V	& 5910	& 1.03	& 200	& 2.1	& 60	& 23.4	& 5.31	& 39.07	& 3	& 22	& Y	& \citealt{moor2006}\\
sig And & 00:18:19.6 +36:47:06&A2V	& 8780	& 2.00	& 155	& 14.9	& -- 	& -- 	& 0.19	& 41.32	& 3	& 22	& Y	& \citealt{morales2009}\\
HD 1466 & 00:18:26.2 -63:28:39&F8V	& 6140	& 1.06	& 140	& 4.7	& -- 	& -- 	& 1.68	& 41.54	& 2	& 22	& Y	& \citealt{smith2006}\\
HD 1461 & 00:18:42.1 -08:03:12&G3V	& 5880	& 1.07	& 65	& 20.5	& -- 	& -- 	& 0.50	& 23.24	& 1	& 22	& Y	& \citealt{lawler2009}\\
9 Cet &  00:22:51.7 -12:12:33&G3V	& 5920	& 0.95	& 40	& 48.5	& -- 	& -- 	& 0.09	& 20.86	& 2	& 60	& Y	& \citealt{chen2014}\\
kap Phe & 00:26:12.3 -43:40:46&A5IVn	& 8010	& 1.76	& 170	& 9.0	& -- 	& -- 	& 0.18	& 23.80	& 3	& 22	& Y	& \citealt{rieke2005}\\
HD 2834 & 00:31:24.9 -48:48:12&A1Va	& 8950	& 2.24	& 95	& 46.3	& -- 	& -- 	& 0.17	& 52.96	& 1	& 22	& Y	& \citealt{chen2014}\\
HD 2772 & 00:31:46.3 +54:31:20&B8Vn	& 12230	& 3.50	& 400	& 7.6	& 35	& 995.0 & 2.29	& 115.74& 4	& 11	& N 	& \citealt{chen2005a}\\
bet03 Tuc & 00:32:44.0 -63:01:53&A0V	& 9050	& 1.63	& 300	& 3.4	& 150	& 13.8	& 2.73	& 45.55	& 5	& 11	& Y	& \citealt{song2001}\\
HD 3126 & 00:34:27.1 -06:30:14&F4V	& 6380	& 1.23	& 200	& 2.9	& 45	& 57.8	& 1.91	& 40.91	& 1	& 22	& Y	& \citealt{trilling2008}\\
HD 3296 & 00:36:02.0 -05:34:15&:F5	& 6450	& 1.46	& 60	& 39.2	& -- 	& -- 	& 0.23	& 45.04	& 1	& 22	& Y	& \citealt{trilling2008}\\
HD 3670 & 00:38:56.7 -52:32:03&F5V	& 6440	& 1.35	& 290	& 1.5	& 55	& 43.1	& 6.47	& 83.12$^{a}$	& 1	& 22	& Y	& \citealt{moor2011}\\
64 Psc & 00:48:58.7 +16:56:26&F8V	& 6570	& 1.52	& 300	& 1.7	& -- 	& -- 	& 1.38	& 23.45	& 2	& 60	& N 	& \citealt{koerner2010}\\
HD 5133 & 00:53:01.1 -30:21:24&K2.5V	& 5170	& 0.66	& 30	& 45.7	& -- 	& -- 	& 0.11	& 14.17	& 3	& 60	& N 	& \citealt{lawler2009}\\
66 Psc & 00:54:35.2 +19:11:18&A1Vn	& 9340	& 2.70	& 250	& 8.7	& 35	& 447.8 & 2.20	& 108.10& 2	& 22	& Y	& \citealt{rhee2007}\\
TYC2278-834-1 & 01:04:25.3 +31:18:27&:K0    	& 5310	& 0.78	& 260	& 0.7	& -- 	& -- 	& 6.19	& 62.89$^{a}$	& 2	& 11	& N 	& --\\
V443 And & 01:10:41.9 +42:55:54&G7V	& 5660	& 0.78	& 200	& 1.4	& 50	& 23.3	& 1.96	& 27.07	& 1	& 22	& Y	& \citealt{kim2005}\\
HR 333 & 01:12:17.3 +79:40:26&A3V	& 8920	& 2.41	& 80	& 69.7	& -- 	& -- 	& 1.88	& 82.84	& 2	& 22	& N 	& \citealt{jura2004}\\
HD 7570 & 01:15:11.1 -45:31:54&F9VFe	& 6170	& 1.19	& 85	& 14.5	& -- 	& -- 	& 0.08	& 15.11	& 2	& 22	& Y	& \citealt{beichman2006a}\\
HD 7590 & 01:16:29.1 +42:56:21&G0V	& 6090	& 0.90	& 200	& 1.9	& 40	& 48.8	& 3.13	& 23.19	& 1	& 22	& Y	& \citealt{plavchan2009}\\
\shortstack{2MASS\\J01203226-1128035}& 01:20:32.3 -11:28:05&G9V	& 5500	& 0.75	& 300	& 0.5	& -- 	& -- 	& 1.96	& 34.39	& 1	& 22	& Y	& \citealt{carpenter2009}\\
HD 8907 & 01:28:34.4 +42:16:02&:F7	& 6250	& 1.23	& 50	& 45.0	& -- 	& -- 	& 2.85	& 34.77	& 1	& 22	& Y	& \citealt{zuckermansong2004a}\\
EO Psc & 01:29:04.9 +21:43:23&K2.5V	& 4930	& 0.96	& 70	& 11.1	& -- 	& -- 	& 0.44	& 23.73	& 1	& 22	& Y	& \citealt{ballering2013}\\
49 Cet & 01:34:37.8 -15:40:34&A1V	& 8670	& 1.77	& 180	& 9.6	& 65	& 73.6	& 11.14	& 59.38	& 8	& 22	& Y	& \citealt{oudmaijer1992}\\
EX Cet & 01:37:35.5 -06:45:38&G5V	& 5440	& 0.75	& 200	& 1.3	& 60	& 14.5	& 1.95	& 23.95	& 2	& 22	& Y	& \citealt{plavchan2009}\\
HD 10472 & 01:40:24.1 -60:59:56&F2IV/V	& 6610	& 1.34	& 200	& 3.4	& 60	& 37.9	& 5.47	& 67.24	& 2	& 22	& Y	& \citealt{zuckermansong2004a}\\
HD 10647$^{b}$ & 01:42:29.5 -53:44:27&F9V	& 6280	& 1.01	& 300	& 1.0	& 55	& 30.9  & 5.80	& 17.43	& 7	& 22	& Y	& \citealt{zuckermansong2004a}\\
\multicolumn{15}{c}{Only a portion of this table has been shown here.  Full content of this table is available in the online material.}
\end{longtable}
\noindent Notes: The spectral types are taken from SIMBAD unless designated by a `:'.  The Num\_Excess parameter describes the number of 
passbands that demonstrate IR excess.  The $\lambda_{start}$ column offers an approximate starting wavelength for the IR excess.  The 
IRS column designates whether this study was able to acquire IRS spectra from the Enhanced Products Archive which in many cases 
supplements the IR excess we report. 
\newline
\noindent Footnotes: 
\begin{list}{$^{a}$}
\item{: This flag indicates the distance shown is from the SED as descrived in detail in the text (Section 3.2.6).}
\end{list}
\begin{list}{$^{b}$}
\item{: This flag indicates that the disk has been resolved through scattered light and plotted in Figure 18.}
\end{list}
\end{landscape}
}

%% file: newreserved.tex
{\footnotesize
\begin{landscape}
\setlength\LTleft{0pt}
\setlength\LTright{4pt}
\setlength\tabcolsep{2pt}
\begin{longtable}{ccccccccccccccc}
\caption{Reserved IR Excess Stars from the Literature and Tycho-2 cross-correlation with AllWISE\label{reserved}}\\

\hline\noalign{\vskip 2mm}
\multicolumn{1}{c}{Name} & \multicolumn{1}{c}{R.A. \& Dec} & \multicolumn{1}{c}{\shortstack{Spectral\\Type}} & \multicolumn{1}{c}{\shortstack{$T_{*,SED}$\\ (K)}}  & \multicolumn{1}{c}{\shortstack{$R_{*,SED}$\\ ($R_{\sun}$)}} & \multicolumn{1}{c}{\shortstack{$T_{dust}$\\ (K)}} & \multicolumn{1}{c}{\shortstack{$R_{disk}$\\ (AU)}} & \multicolumn{1}{c}{\shortstack{$T_{dust2}$\\ (K)}} & \multicolumn{1}{c}{\shortstack{$R_{disk2}$\\ (AU)}} & \multicolumn{1}{c}{\shortstack{$\frac{L_{IR}}{L_{star}}$\\ ($\times10^{-4}$)}} & \multicolumn{1}{c}{\shortstack{Dist.\\ (pc)}} & \multicolumn{1}{c}{\shortstack{Num\\ Excess}} & \multicolumn{1}{c}{\shortstack{$\lambda_{start}$\\($\mu$m)}} & \multicolumn{1}{c}{\shortstack{Has\\ IRS?}} &\multicolumn{1}{c}{\shortstack{Known \\ Reference}} \\
\hline
\endfirsthead

\multicolumn{15}{c}{{\tablename} \thetable{} -- Continued} \\[0.5ex]
\hline\noalign{\vskip 2mm}
\multicolumn{1}{c}{Name} & \multicolumn{1}{c}{R.A. \& Dec} & \multicolumn{1}{c}{Sp. Type} & \multicolumn{1}{c}{$T_{*,SED}$}  & \multicolumn{1}{c}{$R_{*,SED}$} & \multicolumn{1}{c}{$T_{dust}$} & \multicolumn{1}{c}{$R_{disk}$} & \multicolumn{1}{c}{$T_{dust2}$} & \multicolumn{1}{c}{$R_{disk2}$} & \multicolumn{1}{c}{$\frac{L_{IR}}{L_{star}}$} & \multicolumn{1}{c}{Dist.} & \multicolumn{1}{c}{Num} & \multicolumn{1}{c}{$\lambda_{start}$} & \multicolumn{1}{c}{IRS?} & \multicolumn{1}{c}{Known?} \\[0.5ex] \hline
\endhead

\hline
\multicolumn{15}{l}{{Continued on Next Page\ldots}} \\
\endfoot

\\[-1.8ex]
\hline \hline
\endlastfoot
TYC3660-183-1     & 00:00:11.5 +57:52:20 & :A8 	& 7130  &  1.470 & 150 &  7.70 & -- 	  &   --  &  0.910 &132.41 $^{a}$	 &  1  & 22 & N 	& --                   \\    
TYC4026-379-1     & 00:01:18.6 +66:50:12 & :F4 	& 6540  &  1.350 & 260 &  1.90 & -- 	  &   --  &  3.100 &116.38 $^{a}$	 &  2  & 11 & N 	& --                   \\    
TYC2789-507-1     & 00:03:08.3 +42:44:52 & :A2 	& 8430  &  3.080 &  70 &103.90 & -- 	  &   --  &  2.690 &162.33 	 &  2  & 22 & N 	& --                   \\    
TYC4294-584-1     & 00:05:20.0 +68:53:04 & :A1 	& 9130  &  2.410 & 140 & 23.80 & -- 	  &   --  &  0.490 &169.49 	 &  1  & 22 & N 	& --                   \\    
TYC4667-1078-1    & 00:11:34.8 -03:04:38 & :F8 	& 6240  &  1.130 & 180 &  3.10 & -- 	  &   --  &  2.660 &101.46 $^{a}$	 &  1  & 22 & N 	& --                   \\    
HD 870 	          & 00:12:50.0 -57:54:45 & K0V	& 5550  &  0.750 & 300 &  0.60 &   45  & 26.630&  1.570 &20.18 	 &  1  & 60 &    Y  &Lawler et al. 2009    \\   
TYC4026-208-1     & 00:13:03.4 +67:14:46 & :K1 	& 5190  &  0.760 & 170 &  1.60 & -- 	  &   --  & 17.270 &184.32 $^{a}$	 &  2  & 11 & N 	& --                   \\    
HD 987 	          & 00:13:53.2 -74:41:18 & G8V	& 5670  &  0.780 & 125 &  3.70 & -- 	  &   --  &  0.830 &44.40 	 &  1  & 22 & N    &Zuckerman et al. 2011 \\   
HD 1237 	  & 00:16:12.6 -79:51:04 & G8V	& 5650  &  0.820 & 300 &  0.60 & -- 	  &   --  &  1.630 &17.49 	 &  1  & 60 &    Y  &Beichman et al. 2005  \\   
TYC2272-1059-1    & 00:16:42.9 +36:37:47 & :A2 	& 8300  &  2.110 & 140 & 17.20 & -- 	  &   --  &  0.610 &126.41 	 &  1  & 22 & N 	& --                   \\    
39 Psc 	          & 00:17:50.1 +16:19:51 & F6V	& 6400  &  1.180 & 200 &  2.80 & -- 	  &   --  &  0.320 &44.82 	 &  1  & 22 & N 	&Mizusawa et al. 2012  \\   
26 And 	          & 00:18:42.1 +43:47:27 & B8V	& 10850 &   3.760&   75& 169.30& -- 	  &   --  &  2.450 &202.83 	 &  2  & 22 & N 	&Wyatt 2007            \\   
HD 1562 	  & 00:20:00.2 +38:13:35 & G1V	& 5910  &  0.930 &  75 & 13.40 & -- 	  &   --  &  0.580 &24.79 	 &  1  & 60 & N 	&Koerner et al. 2010   \\   
TYC4015-1052-1    & 00:20:53.9 +61:27:42 & :K6 	& 4320  &  0.600 & 120 &  1.80 & -- 	  &   --  &  5.160 &32.11 $^{a}$	 &  1  & 22 & N 	& --                   \\    
TYC8846-897-1     & 00:21:33.4 -66:18:16 & A0V 	& 9330  &  1.720 & 120 & 24.20 & -- 	  &   --  &  0.500 &142.86 	 &  1  & 22 & N 	& --                   \\    
TYC4019-3245-1    & 00:22:05.1 +62:13:13 & :K4 	& 4850  &  0.690 & 165 &  1.30 & -- 	  &   --  &  8.240 &93.56 $^{a}$	 &  1  & 22 & N 	& --                   \\    
TYC1186-730-1     & 00:23:11.9 +20:05:09 & :A8 	& 7190  &  1.230 & 140 &  7.50 & -- 	  &   --  &  1.090 &105.26 	 &  1  & 22 & N 	& --                   \\    
TYC9135-268-1     & 00:34:53.4 -68:35:48 & A0V 	& 6940  &  2.230 & 340 &  2.10 & -- 	  &   --  &  2.300 &182.81 	 &  2  & 11 & N 	& --                   \\    
TYC16-83-1$^{c}$  & 00:37:19.2 +07:29:10 & :M 	& 2830  &  0.130 & 300 &  0.03 & -- 	  &   --  & 47.830 &1.75 $^{a}$	 &  1  & 22 & N 	& --                   \\    
eta Phe 	  & 00:43:21.1 -57:27:47 & A0IV 	& 9080  &  3.760 &  65 &170.80 & -- 	  &   --  &  0.010 &75.52 	 &  1  & 60 & N 	&Su et al. 2006        \\   
HR 189 	          & 00:44:26.1 +47:51:50 & B5V	& 13000 &   3.690&  175&  47.40&   55  &480.070&  5.190 &191.20 	 &  4  & 11 & N 	&Chen et al. 2014      \\   
TYC3667-527-1     & 00:50:59.5 +59:41:35 & :F5 	& 6470  &  1.300 & 180 &  3.90 & -- 	  &   --  &  2.280 &165.10 $^{a}$	 &  1  & 22 & N 	& --                   \\    
TYC8034-360-1     & 00:51:58.7 -49:47:04 & :G2 	& 5900  &  0.960 & 290 &  0.90 & -- 	  &   --  &  7.680 &170.27 $^{a}$	 &  2  & 11 & N 	& --                   \\    
TYC4017-1710-1    & 00:53:28.1 +60:39:56 & A0V 	& 8250  &  3.030 & 140 & 24.40 & -- 	  &   --  &  0.690 &253.16 	 &  1  & 22 & N 	& --                   \\    
HD 5349 	  & 00:55:11.7 -16:58:17 & K0IV	& 5170  &  1.600 & 250 &  1.60 & -- 	  &   --  &  1.180 &48.94 	 &  1  & 22 &    Y  &Chen et al. 2014      \\   
TYC2802-1387-1    & 00:59:26.2 +40:09:18 & :A3 	& 7950  &  1.580 & 150 & 10.30 & -- 	  &   --  &  0.550 &110.37 	 &  1  & 22 & N 	& --                   \\    
TYC3680-352-1     & 01:03:48.4 +58:09:36 & :A3 	& 8300  &  1.580 & 160 &  9.90 & -- 	  &   --  &  1.130 &186.47 $^{a}$	 &  1  & 22 & N 	& --                   \\    
HD 6434 	  & 01:04:40.1 -39:29:17 & G2/3V	& 5920  &  1.090 & 120 &  6.10 & -- 	  &   --  &  0.080 &41.37 	 &  1  & 60 &    Y  &Reid et al. 2007      \\   
TYC4021-1605-1    & 01:05:13.8 +62:25:26 & :K6 	& 4340  &  0.610 & 150 &  1.10 & -- 	  &   --  &  4.240 &36.90 $^{a}$	 &  1  & 22 & N 	& --                   \\    
TYC25-152-1       & 01:05:38.5 +06:38:50 & :G2 	& 5890  &  0.960 & 180 &  2.40 & -- 	  &   --  & 60.130 &259.96 $^{a}$	 &  2  & 11 & N 	& --                   \\    
TYC4021-680-1     & 01:07:35.5 +63:21:11 & :G5 	& 5660  &  0.880 & 300 &  0.70 & -- 	  &   --  &  4.340 &71.15 $^{a}$	 &  2  & 11 & N 	& --                   \\    
\multicolumn{15}{c}{Only a portion of the table has been shown here. Full table available in online materials.} 
\end{longtable}
\noindent Notes: The spectral types are taken from SIMBAD unless designated by a `:'.  The Num\_Excess parameter describes the number of
passbands that demonstrate IR excess.  The $\lambda_{start}$ column offers an approximate starting wavelength for the IR excess.  The
IRS column designates whether this study was able to acquire IRS spectra from the Enhanced Products Archive which in many cases
supplements the IR excess we report.  
\newline
\noindent Footnotes:
\begin{list}{$^{a}$}
\item{: This flag indicates the distance shown is from the SED as descrived in detail in the text (Section 3.2.6).}
\end{list}
\begin{list}{$^{b}$}
\item{: This flag indicates that the disk has been resolved through scattered light and plotted in Figure 18.}
\end{list}
\begin{list}{$^{c}$}
\item{: This star is most likely a giant based on the difference between proper motion magnitude and SED distance. Needs spectroscopic confirmation.}
\end{list}
\end{landscape}
}